\documentclass[preprint2]{aastex63}

\usepackage{natbib}
\usepackage{amsmath}
\usepackage{hyperref}
\usepackage{color}
\usepackage{float}
\usepackage{xspace} 
\usepackage[caption=false]{subfig}

\newcommand{\kepler}{\textsl{Kepler}\xspace}
\newcommand{\gaia}{\textsl{Gaia}\xspace}
\newcommand{\fleck}{\href{https://github.com/bmorris3/fleck}{\texttt{fleck}}\xspace}

\begin{document}

\title{A Relationship Between Stellar Age and Spot Coverage}

\author[0000-0003-2528-3409]{Brett M.~Morris}
\affiliation{Center for Space and Habitability, University of Bern, Gesellschaftsstrasse 6, CH-3012, Bern, Switzerland}T
\email{morrisbrettm@gmail.com}

\begin{abstract}
We investigate starspot distributions consistent with space-based photometry of F, G, and K stars in six stellar associations ranging in age from 10 Myr to 4 Gyr. We show that a simple light curve statistic called the ``smoothed amplitude'' is proportional to stellar age as $t^{-1/2}$, following a Skumanich-like spin-down relation. We marginalize over the unknown stellar inclinations by forward modeling the {\it ensemble} of light curves for direct comparison with the \kepler, K2 and TESS photometry. We sample the posterior distributions for spot coverage with Approximate Bayesian Computation. We find typical spot coverages in the range 1-10\% which decrease with increasing stellar age. The spot coverage is proportional to $t^n$ where $n =-0.37 \pm 0.16$, also statistically consistent with a Skumanich-like $t^{-1/2}$ decay of starspot coverage with age. We apply two techniques to estimate the spot coverage of young exoplanet-hosting stars likely to be targeted for transmission spectroscopy with the James Webb Space Telescope, and estimate the bias in exoplanet radius measurements due to varying starspot coverage.
\vspace{0.5cm}
\end{abstract}

\section{Introduction}

Stars are born rapidly rotating, and dappled with dark starspots in their photospheres \citep{Berdyugina2005}. Starspots are regions of intense magnetic fields which dominate over local convective motions to produce dim, cool regions in stellar photospheres. Starspot coverage shrinks from stellar youth into middle age. Young solar analogues like EK Dra (50 Myr) have hemispheric starspot filling factors in the tens of percent \citep[][]{Strassmeier1998, Jarvinen2018}, while the Sun's hemispheric spot coverage is roughly $0.03\%$ at 4.6 Gyr \citep{Morris2017a}. Many insights into starspots have been learned by analogy from observations of the Sun and its spots \citep{Solanki2003}.

Stellar magnetic activity is perhaps most easily studied via stellar chromospheres, where magnetic active regions shine brighter than the mean photosphere, giving rise to strong emission lines like \ion{Ca}{2} H \& K which correlate with magnetic activity. One of the pivotal observations of stellar magnetic activity was made by \citet{Skumanich1972}, using chromospheric emission line observations from O.C.~Wilson. Skumanich showed that \ion{Ca}{2} H \& K intensities decay as $t^{-1/2}$, and provided observational evidence that stellar rotation is a key feature which drives stellar magnetic dynamos. Many patient observers have carried on studies of stellar chromospheric activity over the last half-century since Skumanich \citep{Baliunas1995, Hall2008}. 

The ability to probe the properties of magnetic active regions in {\it photospheres} has come into focus more recently, due in part to the now widespread availability of space-based photometry from NASA's \kepler, K2 and TESS missions. Space-based photometry is precise enough to measure rotation periods accurately even for relatively inactive stars, enabling photometric detections of flux variations which were previously very difficult or impossible to measure from the ground.

The \kepler mission observed 150,000 stars just above the galactic plane \citep{Borucki2010,Borucki2011}. \citet{McQuillan2014} found that most stars in the \kepler field are consistent with a gyrochronological age of 4.5 Gyr. There was also a young cluster in the original \kepler field, NGC 6811 \citep[1 Gyr;][]{Curtis2019}. Conforming to new hardware constraints, the following K2 mission targeted 400,000 stars in the ecliptic plane \citep{Howell2014}, which measured photometry on stars in several clusters of various ages including Upper Scorpius \citep[10 Myr;][]{Pecaut2016} Praesepe \citep[650 Myr;][]{Douglas2017}, and M67 \citep[4 Gyr;][]{Onehag2011,Barnes2016}. The TESS mission covers 85\% of the sky and collects photometry on the brightest stars \citep{Ricker2014}, including stars in several young associations including the Upper Centaurus Lupus (UCL) association \citep[16 Myr;][]{Pecaut2016}, and the Pisces--Eridanus (Psc-Eri) stream \citep[120 Myr;][]{Curtis2019b}. 

The wealth of precision photometry available for stars of different ages, as well as precise cluster membership catalogs via \gaia observations \citep{GaiaDR2}, makes it possible to investigate how spot coverage varies as stars age. There is a rich history of attempting to invert photometry of active stars to recover stellar surface intensity maps \citep[see review by][]{Lanza2016}. In general these techniques suffer from many degeneracies; a single light curve can be reproduced by a wide variety of spot models. 

In this work, we will make a few critical assumptions to overcome these degeneracies and determine starspot coverages accurately. These assumptions are: (1) that stars of similar age, mass, and rotation period should have similar spot distributions; and (2) the inclination angles of stars are nearly randomly distributed as observed from Earth. If these assumptions are true, then an {\it ensemble} of light curves of stars in a young cluster can be used to constrain their spot distributions. One can imagine that photometric surveys of young clusters are essentially observing the same star at many different inclinations, allowing us to marginalize over the unknown inclinations of the individual stars if we model their light curves as a population. A similar hypothesis was used by \citet{Jackson2013}.

We devise a method for inverting an ensemble of light curves to measure spot coverage as a function of stellar age. In Section~\ref{sec:population}, we describe several samples of stars sourced from \kepler, K2 and TESS photometry. In Section~\ref{sec:fleck}, we outline an efficient algorithm for calculating the rotational modulations of many spotted stars, for comparison with the \kepler, K2 and TESS photometry of young stars. In Section~\ref{sec:abc} we sample the approximate posterior distributions for the spot coverages of stars in each young association using Approximate Bayesian Computation. In Section~\ref{sec:implications} we discuss the implications for stellar dynamos and exoplanet radii.

\section{Stellar Samples} \label{sec:population}

\begin{figure*}
    \centering
    \includegraphics[scale=0.9]{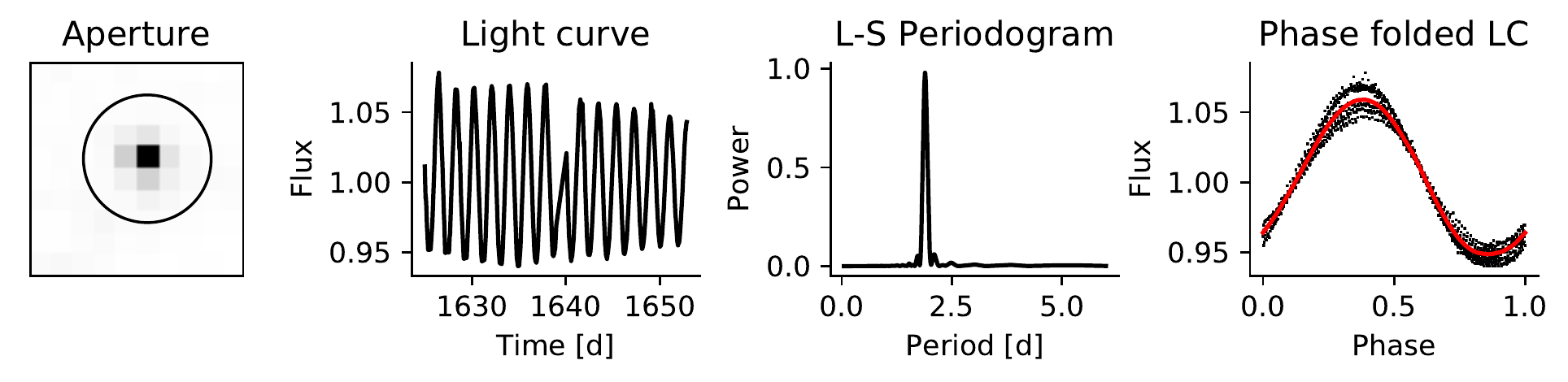}
    \includegraphics[scale=0.9]{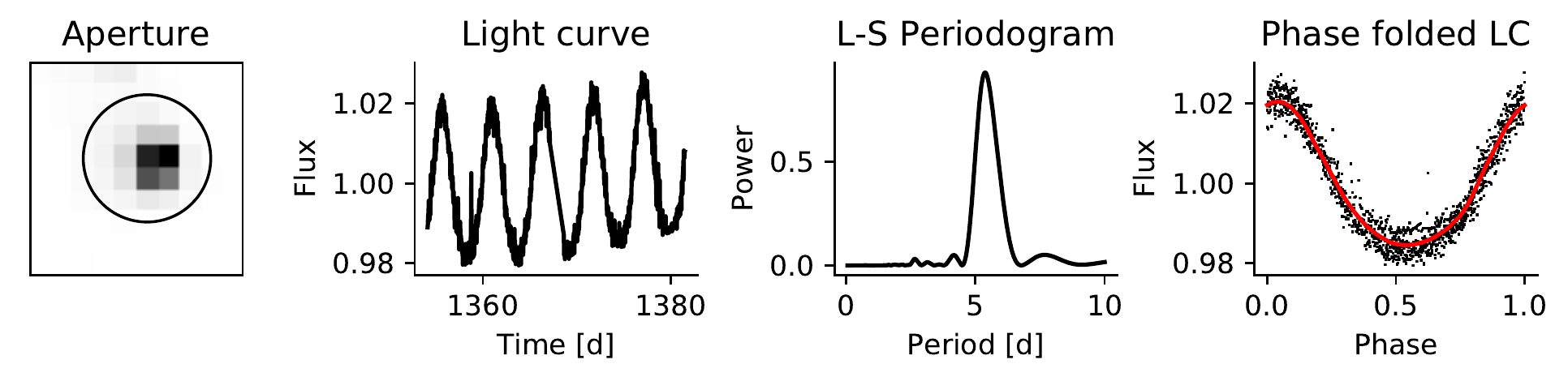}
    \caption{{\it Top row}: TESS Full Frame Image (FFI) photometry of HD 326277, a member of the 16 Myr-old Upper Centaurus Lupus (UCL) association. We measure the flux in a 3 pixel radius circular aperture centered on the stellar coordinates, measure the rotation period with the Lomb-Scargle periodogram, and phase fold the light curve on the rotation period to measure the ``smoothed amplitude'' of the light curve, which is simply the peak-to-trough amplitude of the red curve on the right. {\it Second row}: Same measurements as above for Gaia DR2 4984094970441940864, a member of the Pisces--Eridanus stream (120 Myr). Note that the rotation period is slower and the smoothed amplitude is smaller than HD 326277.}
    \label{fig:photometry}
\end{figure*}

In the present work we limit our analyses to 531 F, G, and K stars. FGK stars likely have fundamentally different dynamos of magnetic activity than M stars, which become fully convective at low masses. For this reason, it may be plausible that FGK stars behave similar to the Sun, whereas M dwarfs quite likely have very different expressions of surface magnetic activity. Also, FGK stars have clearly distinct motions in the observational parameter space which we will explore in this work, while their M-dwarf siblings often have rapid rotation periods even for older clusters. In this analysis, in which we seek to find a relationship between spot coverage and age, we therefore restrict ourselves to the ``solar-type'' FGK stars.

\begin{deluxetable}{lccc}
\tablehead{\colhead{Name} & \colhead{Assoc.} & \colhead{Period} & \colhead{Smoothed}\\ \colhead{} & \colhead{ } & \colhead{[d]} & \colhead{Amplitude}}
\tablecaption{\centering Smoothed amplitudes and rotation periods for all targets considered in this work (full table is available online).}
\startdata
G 132-51 B & Upper Sco & 1.73 & 0.0394 \\
HIP 6276 & Upper Sco & 2.73 & 0.0311 \\
G 269-153 A & Upper Sco & 2.40 & 0.0904 \\
G 269-153 B & Upper Sco & 1.77 & 0.0712 \\
G 269-153 C & Upper Sco & 2.43 & 0.0251 \\
HS Psc & Upper Sco & 3.85 & 0.0666 \\
BD+37 604 Aab & Upper Sco & 4.99 & 0.1422 \\
41 Ari AB & Upper Sco & 10.56 & 0.0737 \\
IS Eri & Upper Sco & 5.39 & 0.1365 \\
HIP 14809 & Upper Sco & 2.23 & 0.0176 \\
HIP 14807 & Upper Sco & 5.33 & 0.0663 \\
V577 Per A & Upper Sco & 2.25 & 0.0377 \\
V577 Per B & Upper Sco & 1.52 & 0.0506 \\
HIP 17695 & Upper Sco & 5.84 & 0.0642 \\
\vdots & \vdots & \vdots & \vdots \\
\enddata
\end{deluxetable}

\subsection{Smoothed amplitudes}

In this section we define what we call the ``smoothed amplitude'' of each light curve. This quantity was first published by \citealt{Douglas2017} for Praesepe members. The smoothed amplitude is the difference between the maximum and minimum flux after the light curve has been phase-folded and smoothed with a Gaussian kernel.

\subsection{Upper Scorpius (USCO): K2}

USCO is a $10 \pm 2$ Myr old part of the nearby Sco-Cen star-forming region \citep{Pecaut2016}. We queried for K2 photometry from FGK members of the young association listed by \citet{Gagne2018}, and found 19 sources.  We measure the stellar rotation period for each star by estimating the peak power in the Lomb-Scargle periodogram \citep{Lomb1976,Press1989}. We then phase-fold each light curve on the best period, smooth the light curve with a Gaussian kernel of width 50-cadences, and report smoothed amplitudes. We visually inspected each light curve for hints of binarity in the periodogram (multiple, non-aliased periods), and discarded any possible binaries and stars with ambiguous rotation periods. 

\subsection{Upper Centaurus Lupus (UCL): TESS}

UCL is a $16 \pm 2$ Myr old part of the nearby Sco-Cen star-forming region \citep{Pecaut2016}. We queried for TESS photometry of sources listed as UCL members by \citet{Gagne2018}, and found 34 sources with TESS Input Catalog (TIC) masses $M > 0.6 M_\odot$ in the full-frame images (FFIs). We query the FFI database for a square region 10 pixels per side, centered on the coordinates of each UCL member. We subtracted the median flux in a 3 pixel radius circular aperture from each FFI, and remove a quadratic trend from each FFI light curve. As in the previous section, for each light curve, we measure the stellar rotation period for each star by estimating the peak power in the Lomb-Scargle periodogram, limiting the maximum period to half of the TESS sector duration. We phase-fold each light curve on the best period, smooth the light curve with a Gaussian kernel of width 50-cadences, and report the smoothed amplitude. Again, we visually inspected each light curve for signs of binarity in the periodogram, and discarded any possible binaries. See Figure~\ref{fig:photometry} for a visual representation of this process.

\subsection{Pisces--Eridanus (Psc-Eri): TESS}

Psc-Eri is a 120 Myr old stellar stream extending 120$^\circ$ across the sky \citep{Meingast2019}. We followed a similar procedure to the previous subsection to produce light curves for each of 100 FGK targets which were identified as members of the Psc-Eri stream by \citet{Curtis2019b}. In addition to using their membership list, we also used the rotation periods reported by \citet{Curtis2019b}, and simply measured the smoothed amplitudes of each light curve after phase folding the light curve and smoothing with a Gaussian kernel with width 100 cadences.

\subsection{Praesepe: K2}


Praesepe is a well-studied, nearby, 650 Myr old cluster. \citet{Douglas2017} measured the amplitudes of rotational modulation of many apparently single stars in Praesepe with K2. Here we will focus on stars that are not classified as binaries or blends. The authors calculated ``smoothed amplitudes'' (which are reported as semi-amplitudes) of the rotational modulation for each star, in which the maximum and minimum flux are measured in smoothed, phase-folded light curves. We adopt the authors' smoothed amplitudes and rotation periods without modification for 220 FGK stars in Praesepe.

\subsection{NGC 6811: \kepler}

NGC 6811 is a 1 Gyr old cluster in the \kepler field \citep{Meibom2011}. \citet{Curtis2019} found what they called a temporary epoch of stalled spin-down for low-mass stars in NGC 6811. We use the \citet{Curtis2019} membership list and rotation periods to build a sample of 167 FGK stars in NGC 6811, and measure smoothed amplitudes from the PDCSAP fluxes for each star in \kepler Quarter 2 using a Gaussian kernel with width 100 cadences. We select only the second quarter of \kepler observations to more closely approximate the variability on timescales similar to the K2 and TESS observations; including the full \kepler light curve tends to average over many spot evolutions and decrease smoothed amplitudes.


\begin{figure}[ht]
    \centering
    \includegraphics[scale=0.83]{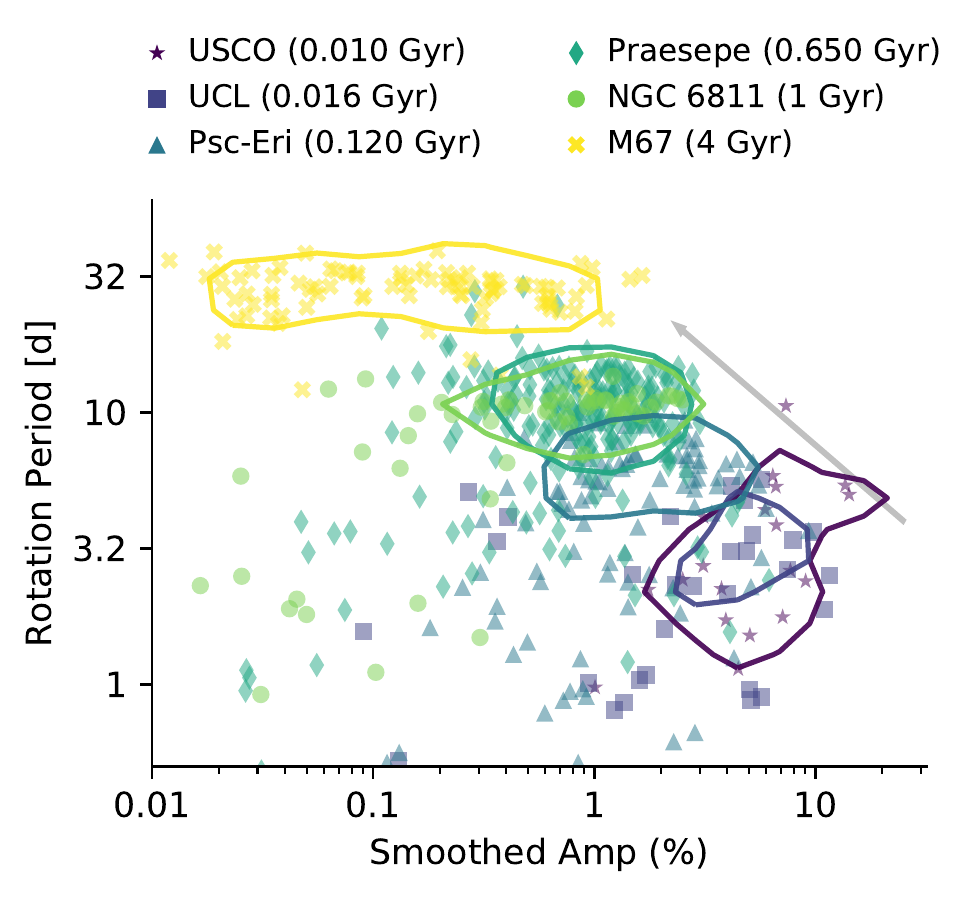}
    \caption{Rotation periods as a function of smoothed amplitudes for the FGK stellar samples defined in Section~\ref{sec:population}. Smoothed contours are drawn around stars in each sample to guide the eye (these contours are drawn by generating a 2D histogram of the stellar samples, smoothing it with a Gaussian, and selecting a cutoff level for drawing the contour). The youngest stars fall in the bottom-right of the plot, corresponding to large longitudinal asymmetries in spot distributions and short rotation periods, while the older stars have smaller smoothed amplitudes, i.e.: more uniform longitudinal starspot distributions, and longer rotation periods. Figure~\ref{fig:age_smap} shows the same observations with an age axis. The spread in the smoothed amplitude axis may be due to: activity cycle phase, stellar inclination, or a combination of the two. }
    \label{fig:prot_smap}
\end{figure}

\begin{figure}[ht]
    \centering
    \includegraphics[scale=0.83]{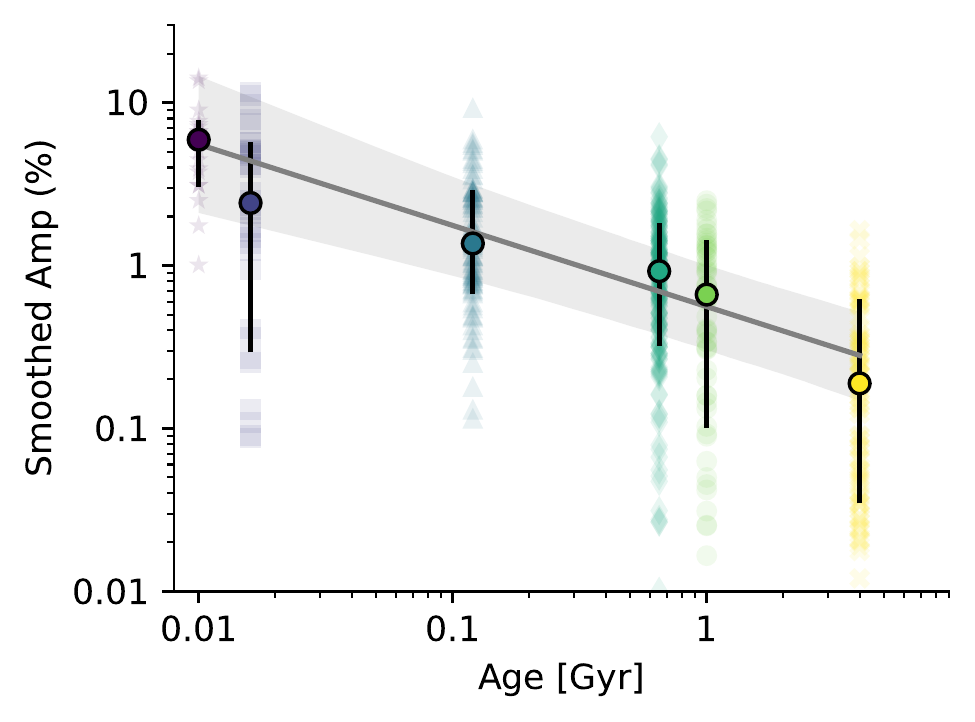}
    \caption{Smoothed amplitudes of the light curves of the FGK stars in each association as a function of age. The colors and symbol shapes correspond to the legend in Figure~\ref{fig:prot_smap}. }
    \label{fig:age_smap}
\end{figure}

\subsection{M67: K2}

M67 is a $4.2 \pm 0.2$ Gyr old cluster in K2 Campaign 5 \citep{Gonzalez2016a, Gonzalez2016b}. We use the cluster membership and rotation periods for 96 stars listed in \citet{Gonzalez2016b} \citep[which are in good agreement with the periods of][]{Barnes2016}, and measure smoothed amplitudes from the PDCSAP fluxes for each star using a Gaussian kernel with width 250 cadences.

\subsection{Trends with rotation period and age}

Figure~\ref{fig:prot_smap} shows a trend in the observable properties of the light curves: there is an anti-correlation between the typical rotation periods of stars in each cluster and the smoothed amplitude of the light curves. One useful perspective encoded in this plot has to do with the {\it axisymmetry} of the starspot distributions. If a star has several starspots which are distributed uniformly in longitude, the rotational modulation amplitude will be relatively small; whereas if spots are concentrated into a small region on one stellar hemisphere, the rotational modulation amplitude will be relatively large. Young stars have short rotation periods and large smoothed amplitudes, corresponding to significant concentrations of dark spots, or significant deviations from uniformly distributed spots. As stars age they drift towards the upper left of the plot (following the direction of the silver arrow); their rotation periods increase and their smoothed amplitudes decrease, or their spots become distributed more uniformly.

Figure~\ref{fig:age_smap} shows another view of the stellar samples in Figure~\ref{fig:prot_smap}, demonstrating the decay of the smoothed amplitude as a function of cluster age. The linear regression trend line (gray) indicates that smoothed amplitudes of light curves generally decline with increasing stellar age, as
\begin{equation}
    \mathrm{Smoothed~Amplitude~[\%]} = \alpha t^m,
\end{equation}
where and $\alpha = {0.56}^{+1.00}_{-0.31}$ and $m = -0.50  \pm 0.17$. 

This power-law index $m$ is remarkably close to the $t^{-1/2}$ decline in chromospheric emission with age discovered by \citet{Skumanich1972}. Perhaps this result suggests there may be a simple relationship between the area in chromospherically active regions and the area in starspots, causing this simple metric for the spot coverage, the smoothed amplitudes, to have the same age dependence as the chromospheric emission index \citep[such relations already exist for magnetic field strength and Ca emission, for example:][]{Schrijver1989}.

\section{Forward modeling ensembles of light curves} \label{sec:fleck}

We now seek to essentially re-calibrate the vertical axis in Figure~\ref{fig:age_smap} by mapping smoothed amplitude distributions onto hemispheric spot covering fractions, $f_S$. In order to do this, we must first devise a technique for simulating photometry of an ensemble of rotating stars.

\subsection{Vectorized ensemble light curve generation}

We simulate ensembles of light curves of stars through a full rotation and measure their smoothed amplitudes using \fleck\footnote{\url{https://github.com/bmorris3/fleck}}. \fleck is a pure Python software package which simulates starspots as circular dark regions on the surfaces of rotating stars, accounting for foreshortening towards the limb, and limb darkening, which is an efficient, vectorized iteration of earlier codes used in \citet{Morris2018b, Morris2019a}. The \fleck algorithm is outlined as follows: suppose we have $N$ stars, each with $M$ starspots, distributed randomly above maximum latitudes ${\rm \ell_{max}}$, observed at $N$ inclinations $\vec{i}_\star$ (one unique inclination per star), observed at $O$ phases throughout a full rotation $\vec{\phi} \sim \mathcal{U}(0, 90^\circ)$. 

We initialize each star such that its rotation axis is aligned with the $\hat{z}$ axis, and set the observer at $x \rightarrow \infty$, viewing down the $\hat{x}$ axis towards the origin. 

We define the rotation matrices about the $\hat{y}$ and $\hat{z}$ axes for a rotation by angle $\theta$:  
\begin{eqnarray}
{\bf R_y}(\theta) &= \begin{bmatrix}
\cos \theta & 0 & \sin \theta \\
0 & 1 & 0 \\
-\sin \theta & 0 & \cos \theta \\
\end{bmatrix} \\
{\bf R_z}(\theta) &= \begin{bmatrix}
\cos \theta &  -\sin \theta & 0 \\
\sin \theta &   \cos \theta & 0\\
0 & 0 & 1\\
\end{bmatrix}
\end{eqnarray}

We begin with the matrix of starspot positions in Cartesian coordinates ${\bf C_i}$, 
\[
{\bf C_i}=
  \begin{bmatrix}
    x_1 & y_1 & z_1 \\
    x_2 & y_2 & z_2 \\
    \vdots  & \vdots  & \vdots \\
    x_M & y_M & z_M
  \end{bmatrix}
\]
for $i=1$ to $N$ with shape $(3, M)$, which we collect into the array ${\bf S}$,  
\[
{\bf S}=
  \begin{bmatrix}
    {\bf C_1}, & {\bf C_2}, & \dots, & {\bf C_N}
  \end{bmatrix}
\]
of shape $(3, M, N)$. We rotate the starspot positions through each angle in $\phi_j$ for $j=1$ to $O$ by multiplying $\bf S$ by the rotation array 
\[
{\bf R_z}=
  \begin{bmatrix}
    [[{\bf R_z}(\phi_1)]], & [[{\bf R_z}(\phi_2)]], & \dots, & [[{\bf R_z}(\phi_O)]]
  \end{bmatrix}
\]
with shape $(O, 1, 1, 3, 3)$. Using Einstein notation, we transform the Cartesian coordinates array $\bf C$ with: 
\begin{equation}
    {\bf R_z}_{[lm\dots]ij} {\bf S}^{j[lm\dots]} = {\bf S^\prime}_{i[lm\dots]}
\end{equation}
to produce a array with shape $(3, O, M, N)$, where $lm…$ indicates an optional additional set of dimensions. Then after each star has been rotated about its rotation axis in $\hat{z}$, we rotate each star about the $\hat{y}$ axis to represent the stellar inclinations $i_{\star, k}$ for $k=1$ to $N$, using the rotation array
\[
{\bf R_y}=
  \begin{bmatrix}
    {\bf R_y}(i_{\star, 1}), & {\bf R_y}(i_{\star, 2}), & \dots, & {\bf R_y}(i_{\star, N})
  \end{bmatrix}
\]
with shape $(N, 3, 3)$, by doing
\begin{equation}
    {\bf R_y}_{[lm\dots]ij} {\bf S^\prime}^{j[lm\dots]} = {\bf S^{\prime\prime}}_{i[lm\dots]}
\end{equation}
which produces another array of shape $(3, O, M, N)$. Now we extract the second and third axes of the first dimension, which correspond to the $y$ and $z$ (sky-plane) coordinates, and compute the radial position of the starspot ${\bf \rho} = \sqrt{y^2 + z^2}$, where $\bf \rho$ has shape $(O, M, N)$. We now mask the array so that any spots with $x < 0$ are masked from further computations, as these spots will not be visible to the observer. We use $\rho$ to compute the quadratic limb darkening 
\begin{equation}
    I(\rho) = \frac{1}{\pi} \frac{1 - u_1 (1 - \mu) - u_2 (1 - \mu)^2}{1 - u_1/3 - u_2/6} 
\end{equation}
for $\mu = \sqrt{1 - \rho^2}$. We compute the flux missing due to starspots of radii $\bf R_{\rm spot}$, which has shape $(M, N)$:
\begin{equation}
{\rm F_{spots}} = \pi {\bf R}_{\rm spot}^2  (1 - c) \frac{I(r)}{I(0)} \sqrt{1 - {\bf \rho}^2}
\end{equation}
The unspotted flux of the star is 
\begin{equation}
    {\rm F_{unspotted}} = \int_0^R 2\pi r I(r) dr,
\end{equation}
so the spotted flux is 
\begin{equation}
    {\rm F_{spotted}} = 1 - \frac{\rm F_{spots,ijk}F_{spots}^{ik}}{\rm F_{unspotted}}
\end{equation}

\begin{figure}
    \centering
    \includegraphics[scale=0.8]{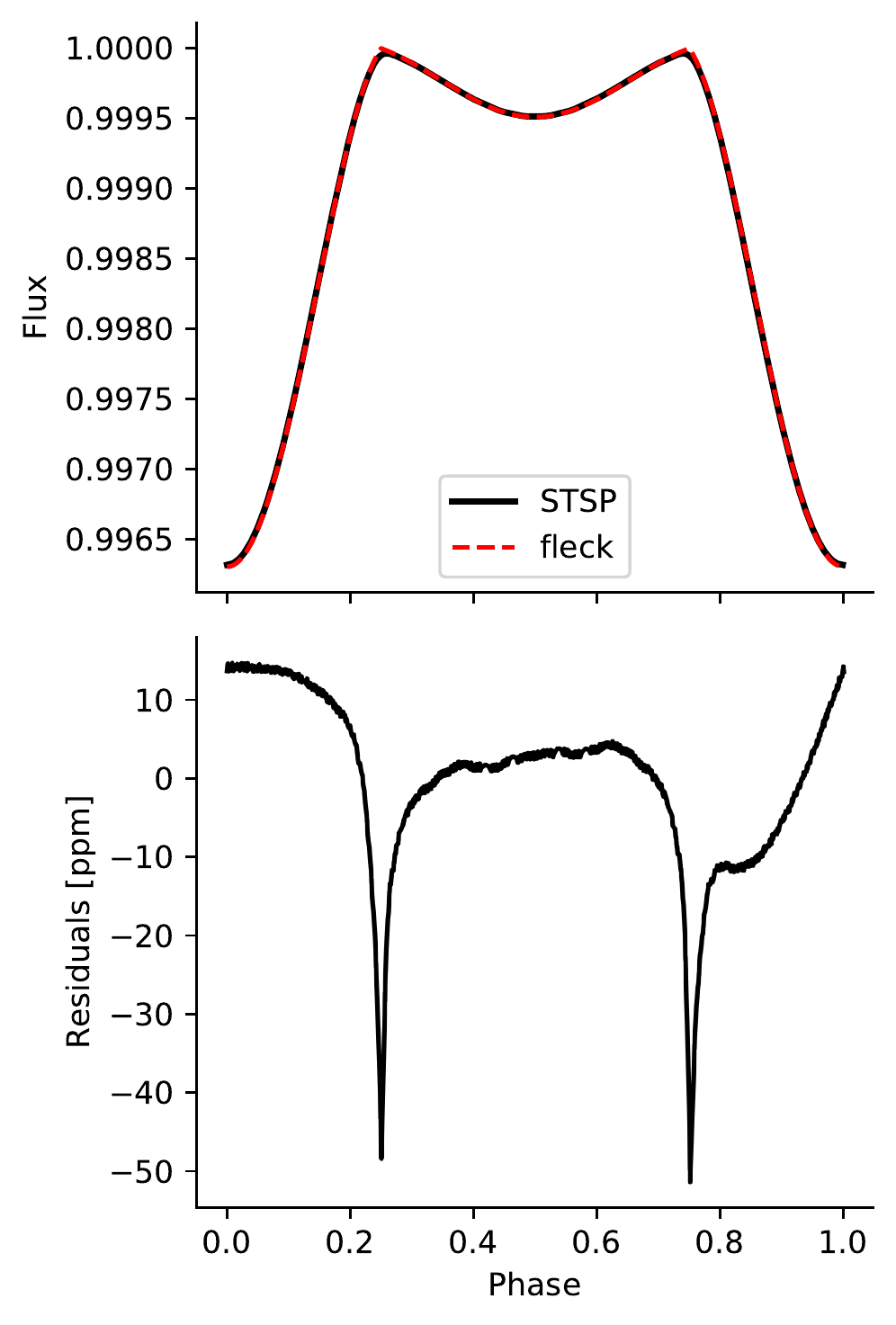}
    \caption{Comparison between the approximate \fleck and the more accurate \texttt{STSP} starspot models, for spots of size $\rm R_{spot}/R_\star = 0.1$. The agreement between models is on the order of the noise in \kepler photometry, so we deem the approximations in \fleck to be valid for the space-based photometry considered here.}
    \label{fig:stsp}
\end{figure}

\begin{figure*}
    \centering
    \includegraphics[width=\textwidth]{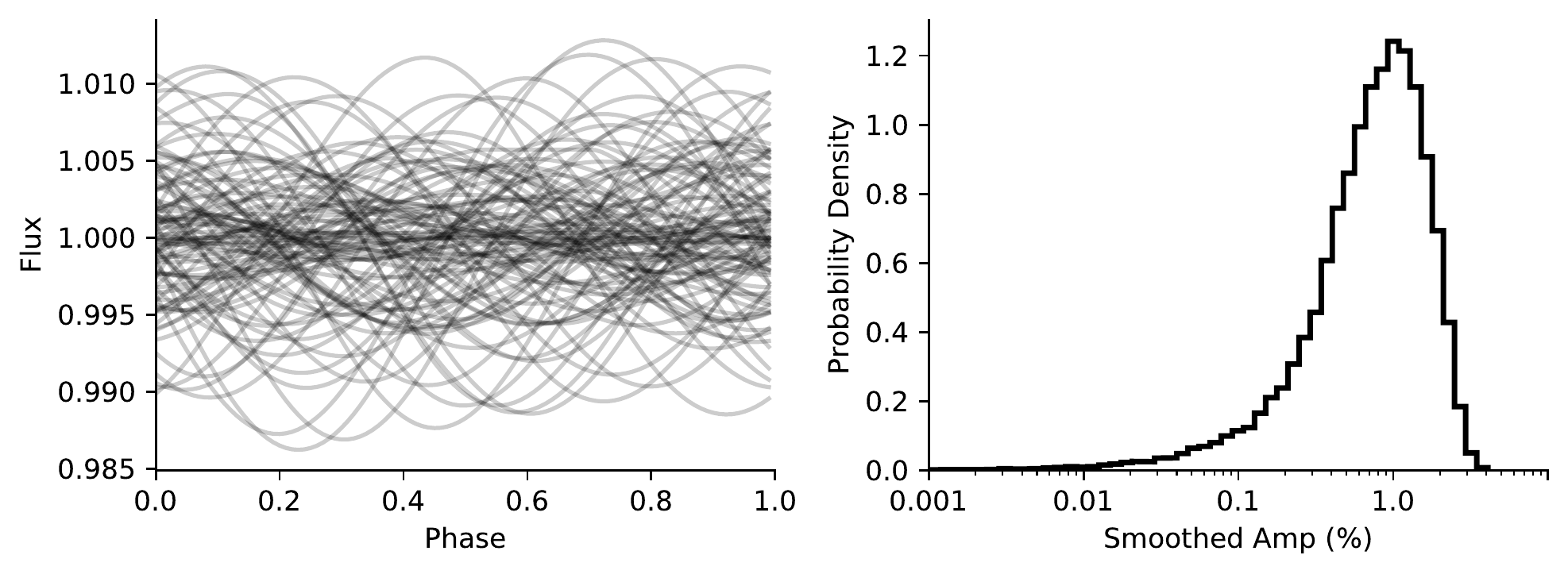}
    \caption{Outputs of the \fleck algorithm, which produces light curves throughout stellar rotations (left), allowing us to measure the difference between the maximum and minimum fluxes for large ensembles of light curves (right). We compare the amplitude distribution of the simulated light curves on the right to the observed distributions of smoothed amplitudes for cluster stars using Approximate Bayesian Computation.}
    \label{fig:fleck}
\end{figure*}


\subsection{Limitations of the model}

The model presented above works best for spots that are small. The array masking step for $x < 0$ does not account for the change in stellar flux due to large starspots which straddle the limb of the star. Large starspots also have differential limb-darkening across their extent, which is not computed by \fleck.  

Comparison with \texttt{STSP}\footnote{\url{https://github.com/lesliehebb/STSP}} is shown in Figure~\ref{fig:stsp}. The models reproduce consistent rotational light curves at the 50 ppm level -- similar to the \kepler noise floor on 1 hour timescales for bright stars \citep{Borucki2011,Christiansen2012}. The maximum divergence between models occurs when the spots are near the limb, where \texttt{STSP} accounts for the spot which straddles the limb and \fleck does not. The differences between the models are small compared to the uncertainties in flux of the K2 photometry, for instance.

\section{Approximate Bayesian Computation} \label{sec:abc}

\begin{figure}
    \centering
    \includegraphics[scale=0.9]{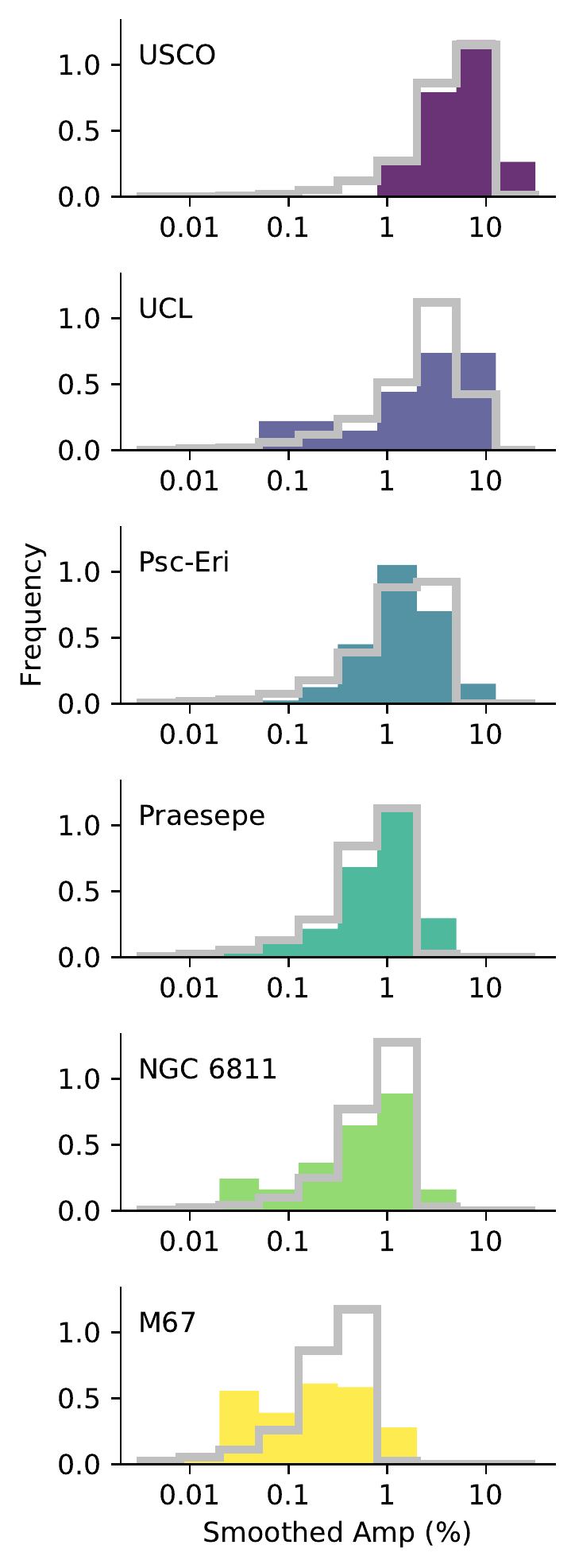}
    \caption{Observed smoothed light curve amplitude distributions (colored histograms) compared with simulated smoothed amplitude distributions (gray) drawn from the posterior distributions for the spot parameters from the ABC technique with \fleck. The goal of the ABC technique is to search for starspot parameters that produce simulated smoothed amplitude distributions which are statistically indistinguishable from the observations.}
    \label{fig:reproduce_smamps}
\end{figure}

\fleck makes it simple to generate new simulated datasets of light curves and their corresponding smoothed amplitude distributions, as shown in  Figure~\ref{fig:fleck}. It is difficult, however, to write down the likelihood of reproducing the observed smoothed amplitudes given a set of spot parameters. When it is straightforward to compute simulated datasets for comparison with observations, but it is difficult to write down the likelihood, Approximate Bayesian Computation (ABC) is a practical tool for sampling from the posterior distributions of parameters \citep{Sunnaker2013,Akeret2015,Dutta2016,Sisson2018}.

In order to find the most likely spot covering fraction $f_S$ given an observed smoothed-amplitude distribution, we explore the spot radius-position-contrast parameter space using ABC. ABC allows us to approximate the posterior PDFs of the spot radius, position and contrast parameters, to ultimately probe which spot covering fractions are compatible with the observations.

The stellar rotational modulation forward-model built with \fleck has three free parameters $\rm \theta^* = \{\ell_{max},\,R_{spot}/R_{star},\,c\}$, the minimum spot latitude above which spots are randomly distributed $\rm\ell_{max}$, the spot radius $\rm R_{spot}/R_{star}$, and the spot contrast $c$ which varies on $[0, 1]$ where $c\rightarrow 0$ approaches perfectly dark spots and $c \rightarrow 1$ are spots with the same intensity as the photosphere. We assign uniform bounded prior probability distributions $\mathcal{U}(0, 90^\circ),\,\mathcal{U}(0, 1),\,\mathcal{U}(0, 1)$, respectively. We use \fleck to generate thousands of light curves of stars observed at random inclinations, producing one trial smoothed-amplitude distribution per $\theta^*$. 

We construct a simple rejection sampling algorithm which operates as follows: (1) perturb the previous step to propose a new set of parameters $\theta^*$ drawn from the prior; (2) use \fleck to compute the smoothed-amplitude distribution for a large sample of stars, (3) for use as a \textit{summary statistic}, we compute the two-sample Anderson-Darling statistic $A^2$ for comparing how close the trial smoothed-amplitude distribution is to the observed one \citep[][see discussion in Appendix~\ref{sec:ad}]{anderson1952,Scholz1987}; (4) if $\rm A^2 < A^2_{crit}$, we accept the proposed step and add it to our chain; (5) go back to step (1), and repeat. We select $\rm A^2_{crit} = 0$, not far from the minimum value of the Anderson-Darling statistic $\rm A^2_{min} \sim -1.3$. 


The approximate posterior distributions produced by ABC should approach the true posterior distributions in the limit that $\rm A^2_{crit} \rightarrow A^2_{min}$, provided that the Anderson-Darling statistic is a \textit{sufficient} statistic. In practice it is difficult to prove that a statistic is sufficient, so we note that the posterior distributions shown here are valid given the hypothesis that the Anderson-Darling statistic is a sufficient one.

The posterior distributions from the ABC analysis illustrate the three-way degeneracy between starspot latitudes, radii and contrasts for stars with unknown inclinations. For a fixed number of spots, small spots spread randomly over all latitudes generate rotational modulations similar to larger spots concentrated near the poles. Similarly, small spots with extreme intensity contrasts ($c\rightarrow 0$) reproduce similar rotational modulations to larger spots with less extreme spot contrasts ($c\rightarrow1$). This exercise is a demonstration of why it is so difficult to invert light curves of rotational modulation and recover unique starspot properties -- a wide range of spot parameters can produce similar light curves.

Figure~\ref{fig:reproduce_smamps} shows the simulated smoothed amplitude distributions (gray histograms) with spot parameters drawn randomly from the posterior distributions, compared with the observed smoothed amplitude distribution for stars in each association (colored histograms). This figure illustrates how the ABC algorithm minimized the Anderson-Darling statistic between the simulated and observed smoothed amplitude distributions. Most simulated and observed distributions are statistically indistinguishable according to the Anderson-Darling statistic. 

The poorest ``fit'' is M67, the oldest cluster, for which the simulations produce a more strongly peaked smoothed amplitude distribution near 0.5\%. The lack of a peak in the observed smoothed amplitude distribution for M67 is likely real; observational bias in favor of detecting large amplitude variability would cause a peak at large smoothed amplitudes. The lack of a peak in the smoothed amplitude distribution may be a hint that the spin axes of the stars in M67 are not randomly distributed (see further discussion in Section~\ref{sec:spins}).

\begin{figure*}
    \centering
    \includegraphics[scale=0.82]{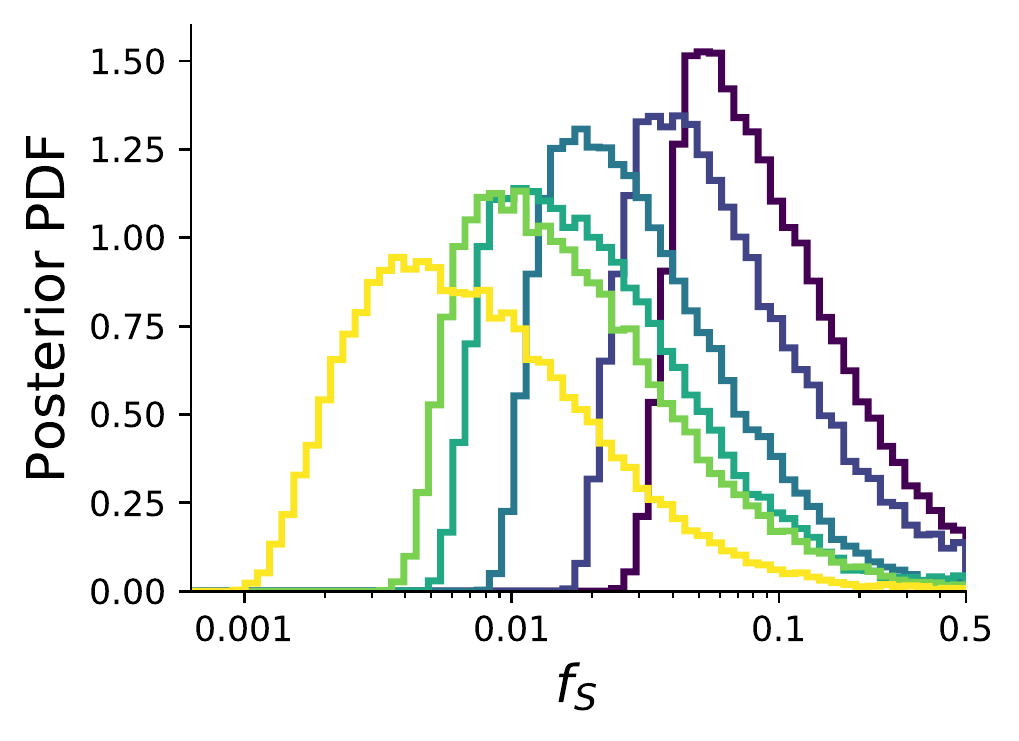}
    \includegraphics[scale=0.8]{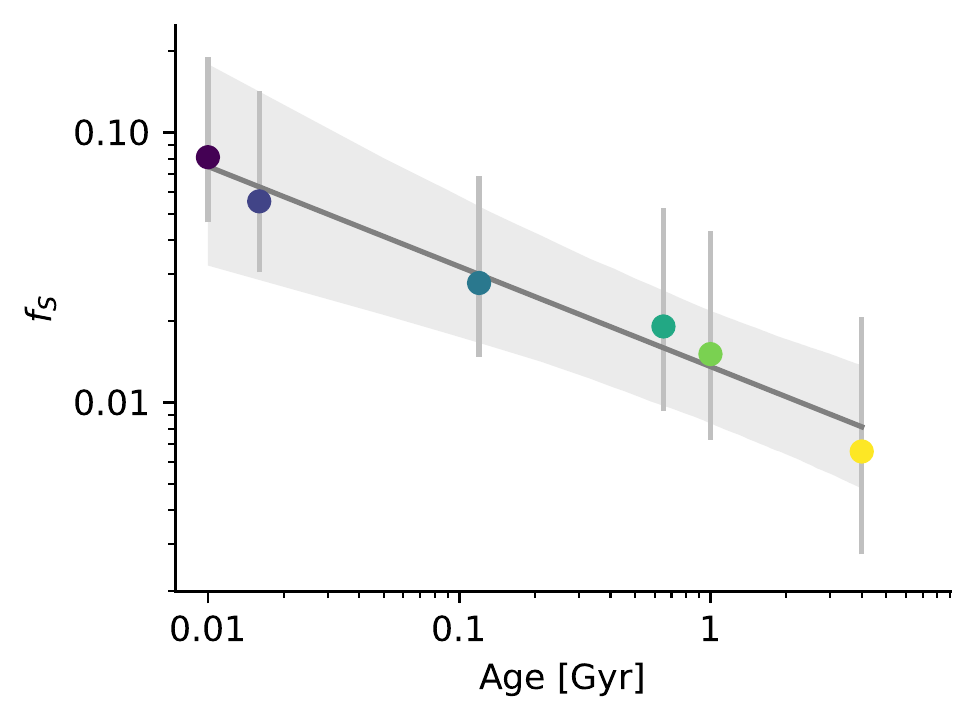}
    \caption{Approximate posterior distributions for the total spot coverage $f_S$ for stars in each association. Colors correspond to the legend in Figure~\ref{fig:prot_smap}. As stars age (from purple through yellow), the most likely spot coverage $f_S$ decreases from 8\% at 10 Myr (USCO, purple) down to 0.8\% at 4 Gyr (M67, yellow). Colors correspond to the legend in Figure~\ref{fig:prot_smap}. 
    }
    \label{fig:fs}
\end{figure*}

The approximate posterior distributions for the total spot coverage is shown in Figure~\ref{fig:fs}. The ABC technique constrains the spot coverage for each stellar sample to between $0.05 < f_S < 0.2$ for the youngest stars (at 10 Myr in USCO), and $0.002 < f_S < 0.02$ for the oldest stars (at 4 Gyr in M67). 

\section{Discussion}\label{sec:implications}

\subsection{A relation between starspot coverage and stellar age} \label{sec:relation}

The relationship between starspot coverage and age can be deduced from the slope and intercept in the right panel of Figure~\ref{fig:fs}. We find the hemispheric starspot covering fraction $f_S$ is related to the stellar age $t$ in Gyr by the simple relation
\begin{equation}
    f_S \approx a t^n \label{eqn:fst}
\end{equation}
where $a = {0.014}^{+0.022}_{-0.008}$ and $n =-0.37 \pm 0.16$. The power law index $n$ is statistically consistent with the approximate inverse square root relation between chromospheric activity and stellar age by \citet{Skumanich1972}. 

One must take care not to infer upper or lower limits to the spotted coverage of individual stars of a given age from Equation~\ref{eqn:fst}. The light curve ensemble modeling technique constrains a {\it typical} spot coverage for stars in each sample, and outliers are likely to exist which will not fall neatly within the confidence intervals of Equation~\ref{eqn:fst}. 

LkCa 4, for example, is a weak-lined T Tauri star in the Taurus Molecular Cloud \citep[$<10$ Myr;][]{Kenyon1995}. High-resolution near-infrared IGRINS spectra from \citet{Gully2017} constrained the star's spot coverage to $f_S \sim 0.8$. The authors argue that such a large coverage by dark regions challenges our notion of ``spot coverage,'' since the majority of the stellar photosphere emits at a cooler temperature than its spectral type \citep{Pettersen1992}. The ensemble light curve modeling technique assumed three starspots which cover a {\it minority} of the stellar surface, so the results (Equation~\ref{eqn:fst}) should not be applied to stars covered in mostly cool regions, like LkCa 4. 



\subsubsection{Relating smoothed amplitude to spot coverage}

Given that we have established a relationship between spot coverage and age and smoothed amplitude and age, we can infer the relationship between spot coverage and smoothed amplitudes, 
\begin{equation}
    f_S = a \left(\frac{\mathrm{Sm~Amp~[\%])}}{\alpha}\right)^{n/m},
\end{equation}
or approximately
\begin{equation}
    f_S \sim 0.02 \times \left(\mathrm{Sm~Amp~[\%]}\right)^{0.74}.
\end{equation}
We caution users of this formula that it only describes how ensembles of stars behave on average, and that one should not infer spot coverage for individual stars directly from light curve amplitudes, as the light curve amplitude is degenerate with the stellar inclination and spot contrast.



\subsection{Second-order effects}

In this subsection we discuss several factors which may have small but significant affects on the conclusions drawn from the ABC analysis. 

\subsubsection{Comparing rotational modulation across bandpasses}

One difficulty in comparing photometry across two telescopes is that \kepler, TESS, and Gaia all have a slightly different bandpasses. Fortunately, the effect of the slightly different bandpasses on the scale of rotational modulation is small \citep[see Figure 2 of][]{Morris2018b}. Future catalogs of photometry from the \gaia mission may also prove useful in measuring the photometric variability of young stars due to starspots.

\subsubsection{Activity cycles}

Magnetic activity cycles with timescales of years to decades will be a source of imprecision in the spot distribution analysis. Stars observed near activity minima will have smaller smoothed amplitudes than identical stars observed near activity maxima, creating a distribution in smoothed amplitude space for even a single star observed through time. Furthermore, the properties of the activity cycle vary with stellar age; \citet{Baliunas1995} found that younger, rapidly-rotating G and K stars have more stochastic variations when compared with older, more slowly rotating stars with smooth, cyclic activity patterns. 

In this work we assume that by observing samples of tens to hundreds of stars at each age, we are observing stars at a variety of activity cycle phases. In this sense, we may be implicitly marginalizing over the latent activity cycle phase variable within each stellar subsample. Some stars will be observed near minimum and have smaller light curve amplitudes, while others will be observed near maximum and have larger amplitudes. The broad confidence interval shown in Figure~\ref{fig:fs} may therefore already account for some of the apparent broadening in the $f_S$ distributions of stars due to activity cycles.

\subsubsection{Metallicity and magnetic activity cycles}

It has been claimed that metallicity may affect the photometric variability of stars throughout their magnetic activity cycles. \citet{Karoff2018} measured the photometric variations of HD 173701, a star with twice the solar metallicity, and found that its variability amplitude is significantly larger than solar. When more photometry and cluster membership catalogs become available, it may be necessary to add a third dimension to the analysis of spot coverage as a function of stellar age, which parameterizes variation in spot coverage with stellar metallicity.

\subsubsection{Starspot evolution}

\citet{Giles2017} showed that starspots have longer lifetimes on cooler stars. When one phase folds the light curve of a solar-mass star, the light curve amplitude may vary strongly as a function of the duration of the bin over which the light curve is phase folded. Smaller duration bins will match the comparatively short lifetimes of starspots on Sun-like stars ($\lesssim$ months), giving accurate representations of the true light curve amplitude. Larger duration bins, like the four-year \kepler light curves of NGC 6811, will integrate over several spot evolutions and therefore may dilute the true amplitude of the light curve variation. For this reason, we used a single \kepler quarter rather than the full \kepler light curves for NGC 6811.

\subsubsection{Stellar inclination distribution} \label{sec:spins}

We have assumed that spin axes of stars are distributed randomly. Spins of stars in old open clusters may be preferentially aligned with one another \citep{Corsaro2017, Kovacs2018}, which would introduce yet another set of degenerate parameters into the ABC analysis. 




\subsection{Comparison with observations of planet-hosting stars}

\begin{deluxetable*}{lccccc}
\tablecaption{Spot coverage on planet-hosting stars\label{tab:pred_vs_measured}}
\tablehead{\colhead{Star} & \colhead{Spectral} & \colhead{Photometry} & \colhead{Age}    &  \multicolumn{2}{c}{$f_S$}\\
& \colhead{Type}     & \colhead{Source}     & \colhead{[Myr]}  &  \colhead{Predicted} & \colhead{Measured} }
\startdata
V1298 Tau & K0  & K2       &$23 \pm 4^a$ & ${0.05}^{+0.06}_{-0.02}$ & ${0.09}^{+0.01}_{-0.02}$ \\
DS Tuc A  & G6V & TESS     &$45 \pm 4^b$ & ${0.04}^{+0.04}_{-0.02}$ & ${0.071}^{+0.003}_{-0.003}$ \\
Qatar-4   & K1V & TESS     &$170 \pm 10^c$ & ${0.03}^{+0.02}_{-0.01}$ & ${0.030}^{+0.009}_{-0.006}$ \\
Kepler-411& K2V & \kepler  &$212 \pm 31^d$ & ${0.02}^{+0.02}_{-0.01}$ & ${0.017}^{+0.003}_{-0.002}$ \\
WASP-52   & K2V & HST/WFC3 &$400^{+300~e}_{-200}$ & $0.01-0.04$ & $0.05 \pm 0.01^h$ \\
Kepler-289& G0V & \kepler  &$650 \pm 440^f$  & ${0.015}^{+0.009}_{-0.006}$ & ${0.031}^{+0.002}_{-0.005}$ \\
EPIC 247589423 & K5.5 & K2 & $687 \pm 063^j$ & ${0.014}^{+0.009}_{-0.006}$ & ${0.0066}^{+0.0017}_{-0.0016}$ \\
K2-100    & G0V  & K2  & $790 \pm 30^k$ & ${0.014}^{+0.009}_{-0.006}$ & ${0.032}^{+0.003}_{-0.001}$ \\
K2-101    & K2V  & K2  & $790 \pm 30^k$ & ${0.014}^{+0.009}_{-0.006}$ & ${0.035}^{+0.0004}_{-0.003}$ \\
Kepler-21 & F6V & \kepler & $2600\pm 160^l$  & ${0.009}^{+0.006}_{-0.003}$ & $<0.001$   \\
Kepler-50 & F7V  & \kepler & $3590^{+780~l}_{-450}$ & ${0.007}^{+0.005}_{-0.003}$ &  $<0.001$ \\
Sun       & G2V & ---      & $4570 \pm 10^g$& ${0.007}^{+0.005}_{-0.003}$ & $<0.005^i$
\enddata
\tablecomments{Spot coverages $f_S$ ``predicted'' from the stellar ages (via Equation~\ref{eqn:fst}) compared with ``measured'' spot coverages from direct modeling of the light curves with \fleck (for all targets except WASP-52), and HST/WFC3 spot occultation measurements (for WASP-52). Predicted and measured spot coverages are plotted in Figure~\ref{fig:fs_comparison}. }
\tablerefs{(a) \citet{David2019, David2019b}; (b) \citet{Newton2019}; (c) \citet{Alsubai2017}; (d) \citet{Sun2019}; (e) \citet{Hebrard2013, Kirk2016, Bruno2018}; (f) \citet{Schmitt2014};  (g) \citet{Sonett1991}; (h) \citet{Bruno2019}; (i) \citet{Morris2017a}; (j) \citet{Ciardi2018,Mann2017b}, (k) \citet{Mann2017}; (l) \citet{Silva2015}}
\end{deluxetable*}

\begin{figure*}[!ht]
    \centering
    \includegraphics[scale=1]{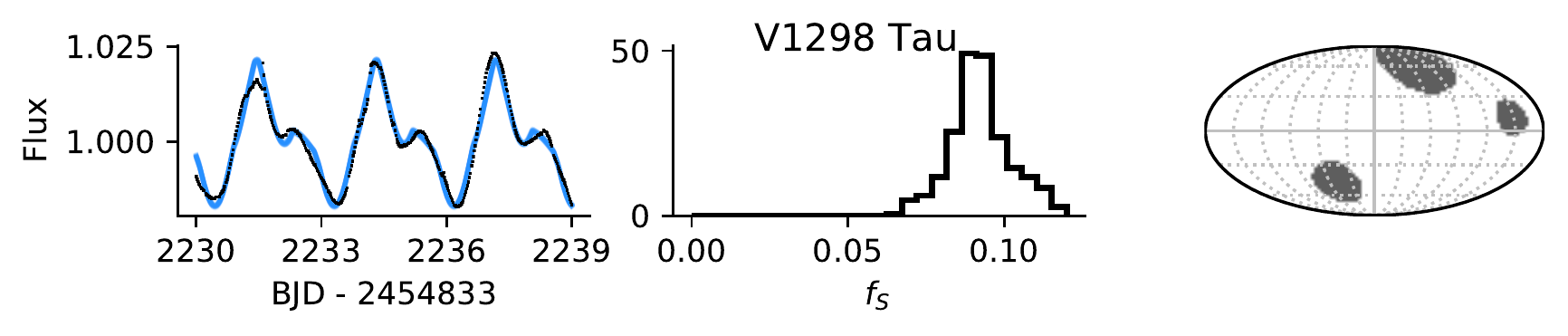}
    \includegraphics[scale=1]{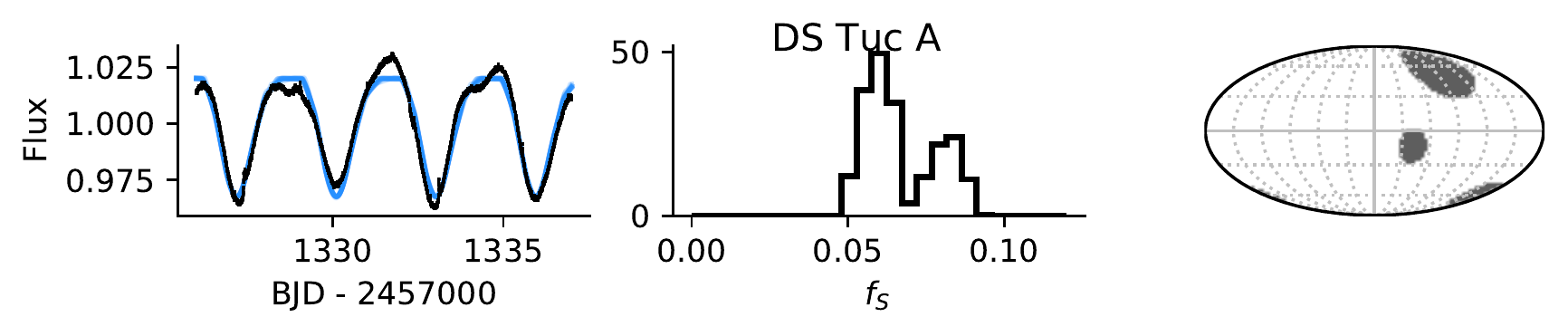}
    \includegraphics[scale=1]{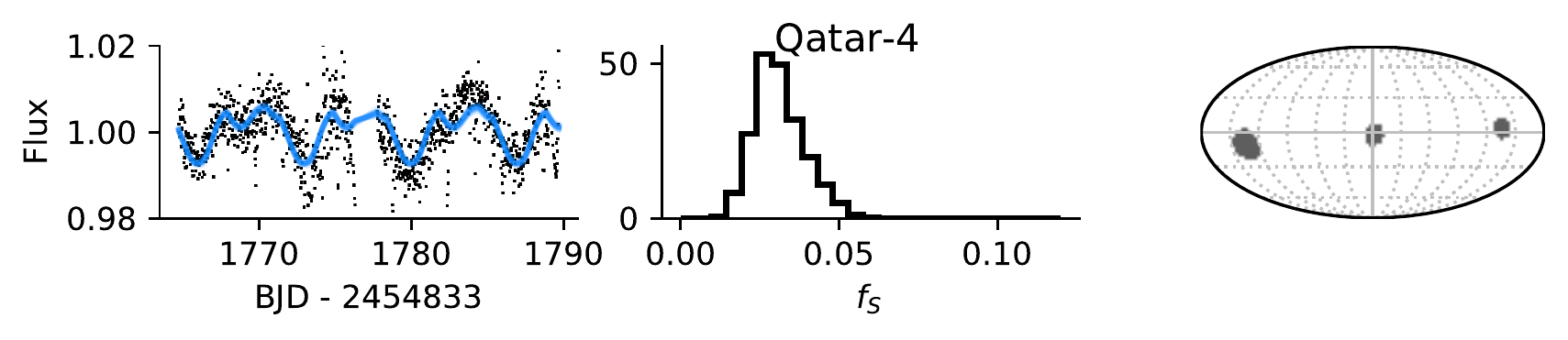}
    \includegraphics[scale=1]{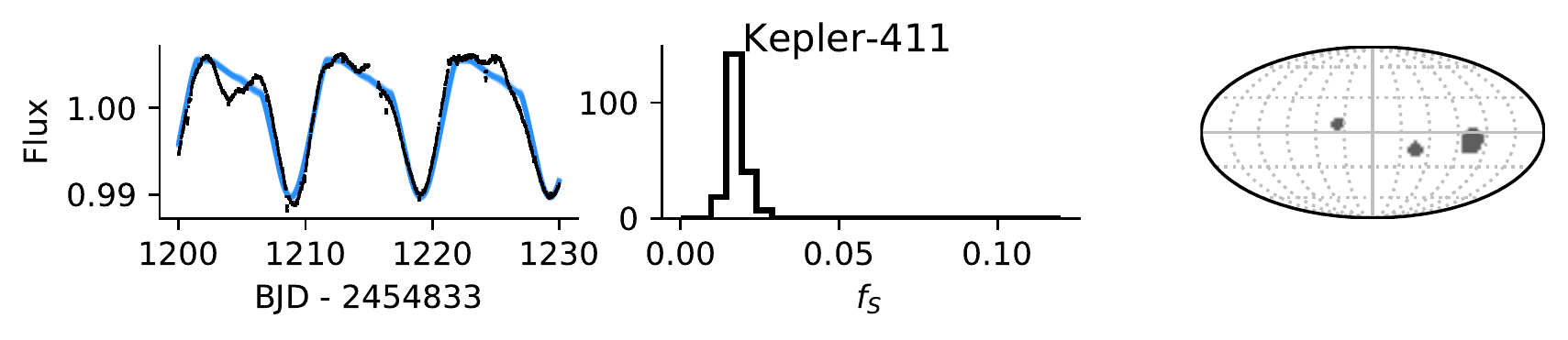}
    \includegraphics[scale=1]{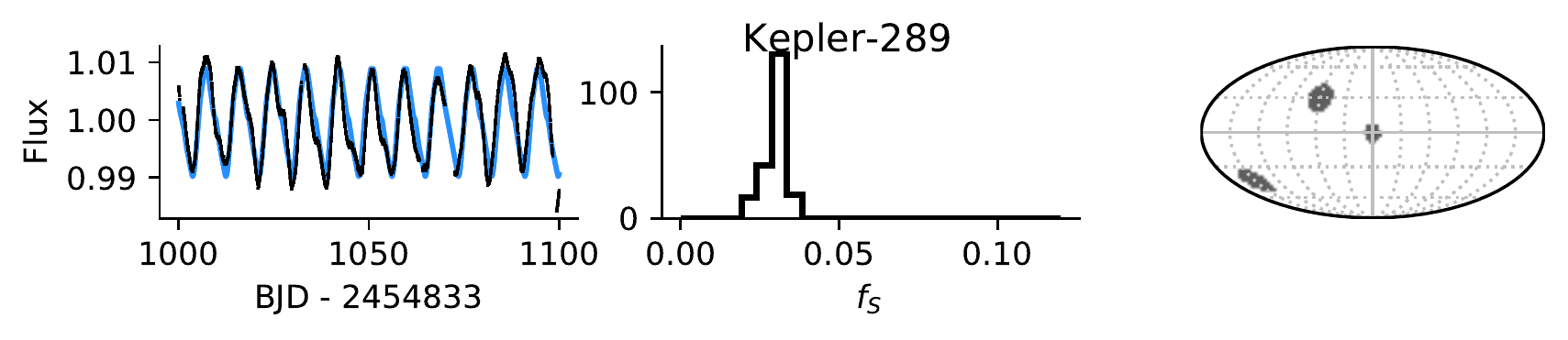}
    \caption{Starspot rotational modulation models genereted with \fleck (left, blue) for selected portions of observations (left, black); the posterior distributions for the spot coverage (middle); and a single draw from the spot parameter posterior distributions visualized in the Mollweide projection (right). Note that the solutions to the spot latitudes are perfectly degenerate about the equator. Figure continues on next page.}
    \label{fig:bespoke}
\end{figure*}

\begin{figure*}[!ht]
    \ContinuedFloat
    \centering
    \includegraphics[scale=1]{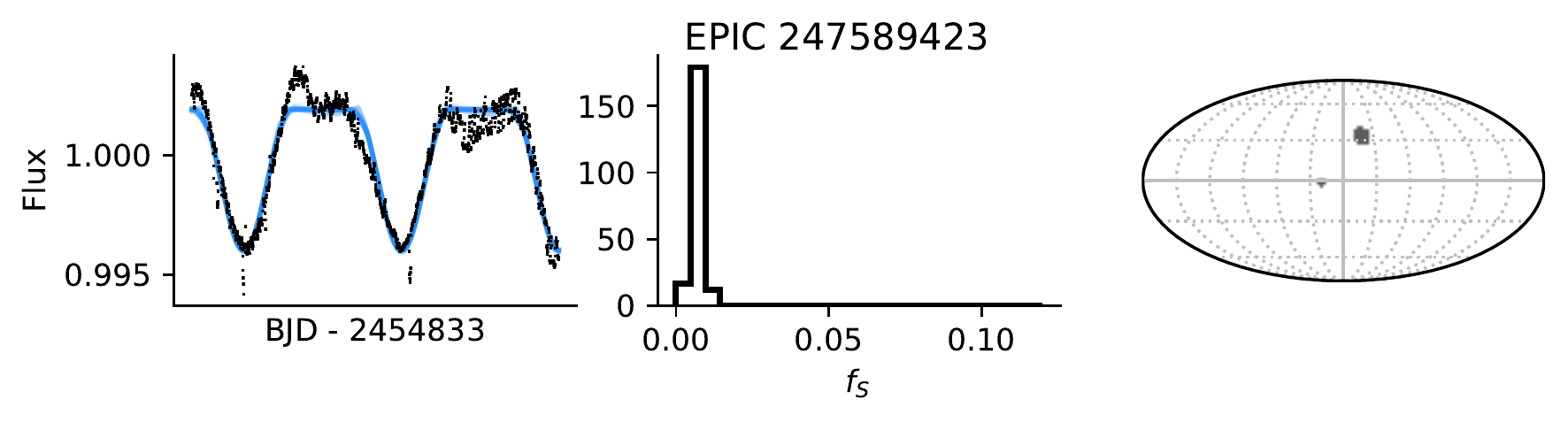}
    \includegraphics[scale=1]{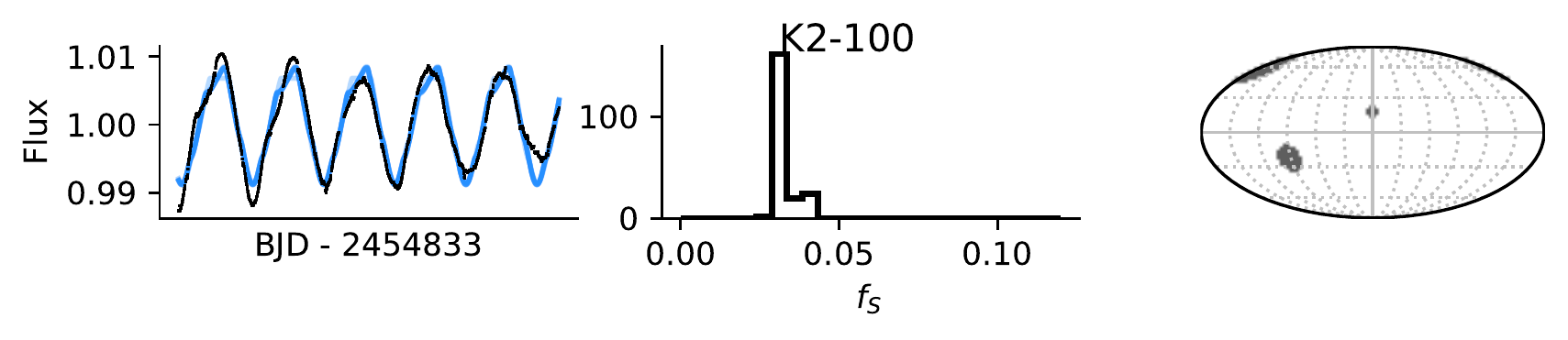}
    \includegraphics[scale=1]{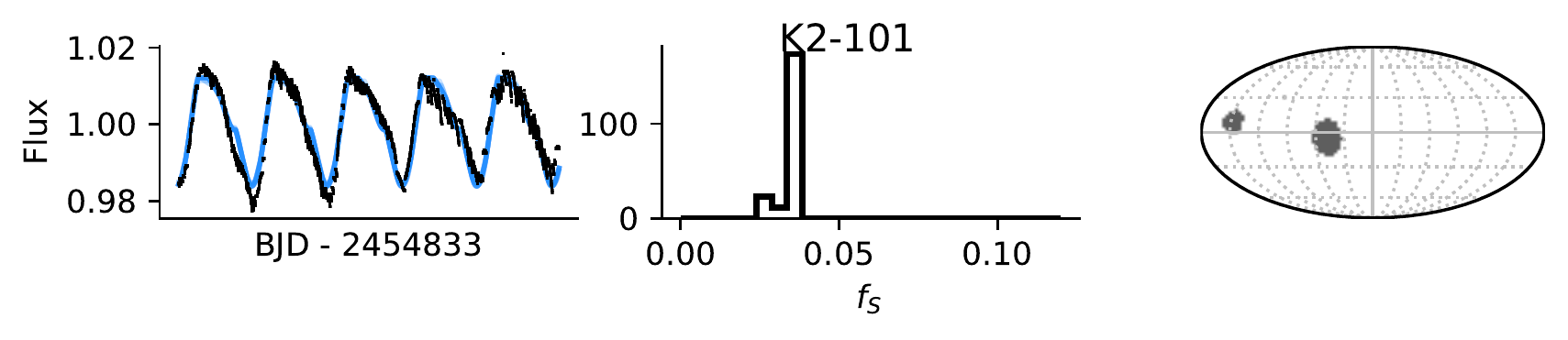}
    \includegraphics[scale=1]{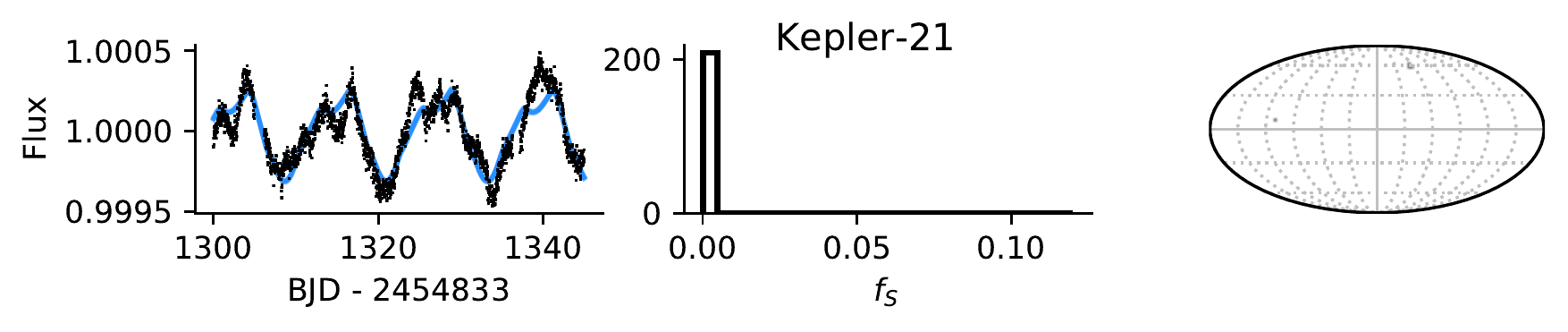}
    \includegraphics[scale=1]{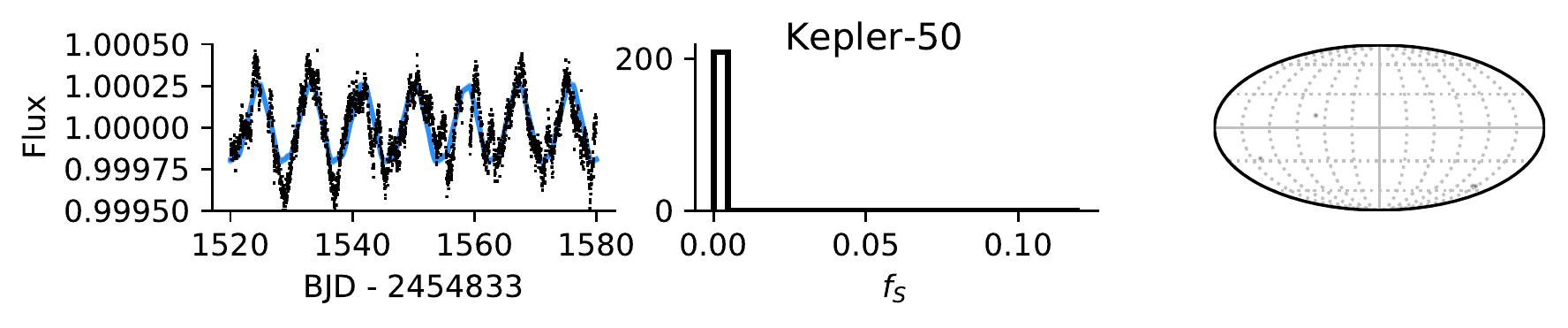}
    \caption[]{(continued)}
\end{figure*}

In addition to ensemble light curve modeling, we can also use \fleck to directly model the photometry of individual stars. In this section, we fit the \fleck model to the light curves of spotted stars with Markov Chain Monte Carlo (MCMC) to validate the spot coverage relation in Equation~\ref{eqn:fst} for stars with precise photometry and ages, listed in Table~\ref{tab:pred_vs_measured}.

As discussed earlier, the positions, radii, and contrasts of starspots are degenerate with one another, making it difficult to extract precise spot properties from the rotational modulation of planet hosting stars. Thus we make several simplifying assumptions that allow us to fit for the spot coverage on these stars. First, we assume that the stellar inclination is $i_\star = 90^\circ$, which may be a good approximation since each of these systems host (often multiple) transiting exoplanets. Next, we fix the spot contrast to $c=0.7$; this is compatible with the area-weighted spot coverage of sunspots, the starspot contrasts of HAT-P-11 \citep{Morris2017a}, and a valid approximation to the observed spot contrasts of several stars extrapolated into the \kepler and TESS bandpasses \citep{Morris2018b}. We also place a uniform bounded prior on the spot latitudes $|\ell| < 60^\circ$. It is necessary to impose this latitude prior because without it the Markov chains occasionally prefer spots with implausibly large radii, skewing the fits towards large spot coverage; this prior is consistent with by solar observations since sunspots are not observed above $\sim 45^\circ$ \citep{Morris2019a}. Finally, we also enforce the prior that $\rm 0 < R_{spot}/R_{star} < 1$, ensuring that the largest spots are still small compared to the stellar radius, to forbid spots with radii several orders of magnitude larger than the largest sunspot groups.

We choose to fix the number of starspots to three in each fit. We find that fitting less than three spots does not accurately reproduce the observed rotational modulation, and fitting additional spots does not significantly improve the fits. 

\begin{figure}
    \centering
    \includegraphics[scale=0.83]{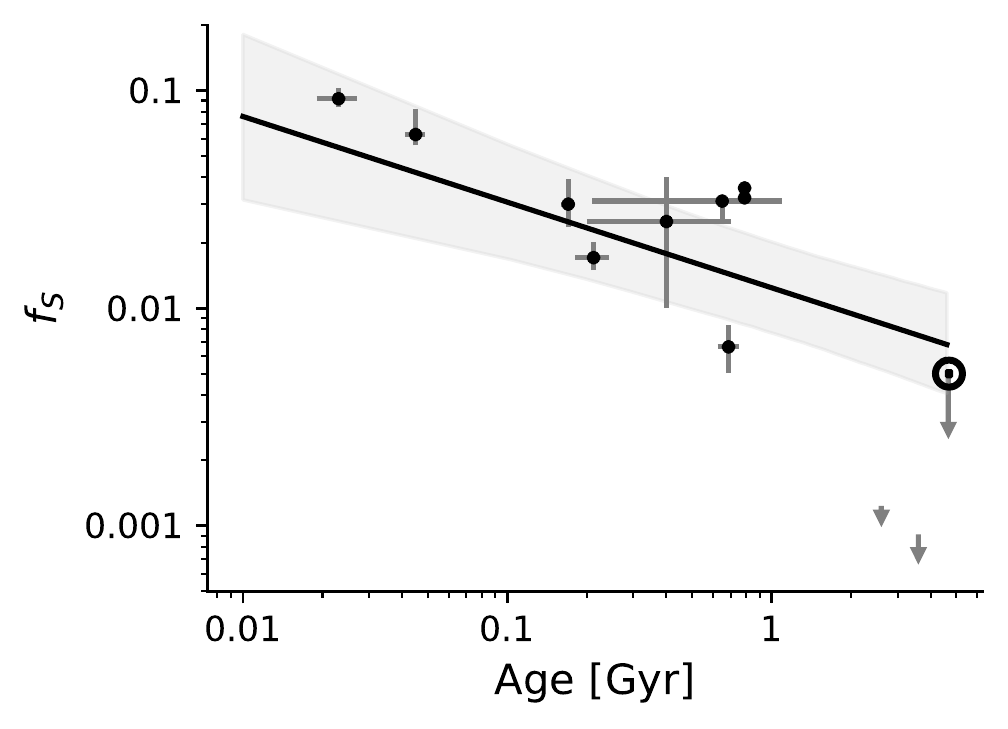}
    \caption{Comparing the posterior spot coverage distributions from direct modeling of light curves of planet hosting stars in Table~\ref{tab:pred_vs_measured} with the relation in Equation~\ref{eqn:fst} (black curve, grey $1\sigma$ confidence interval). The solar measurement of spot coverage is an upper limit based on the most-spotted observation of the Sun.}
    \label{fig:fs_comparison}
\end{figure}

The photometry, maximum-likelihood models, spot coverage posterior distributions, and spot maps are shown in Figure~\ref{fig:bespoke}. The three-spot model reasonably approximates the rotational modulation in each light curve. Then we compare the posterior distributions for the expected spot coverage inferred from the \fleck models with the spot coverage estimate from Equation~\ref{eqn:fst} in Figure~\ref{fig:fs_comparison}. Most stars fall within the $1\sigma$ confidence interval for the predicted spot coverage. 

We include an upper-limit for the spot coverage of the Sun in Figure~\ref{fig:fs_comparison} using the Mount Wilson Observatory sunspot catalog \citep{Howard1984}, analyzed in the stellar context by \citet{Morris2017a}. 

Equation~\ref{eqn:fst} tends to over-predict the amplitudes of rotational modulation in the oldest stars: Kepler-21, Kepler-50 and the Sun (Figure~\ref{fig:fs_comparison}). Some possible explanations for overestimate from include: (1) cluster stars like those of M67, which are the anchors for the spot coverage relation near 4 Gyr, are more active than field stars like Kepler-21 and Kepler-50; (2) spots are more uniformly distributed on old field stars, diminishing the rotational modulation amplitude; (3) the field stars could have been observed near activity minimum; or (4) we may be viewing Kepler-21 and Kepler-50 at low inclinations, producing small light curve amplitudes. A larger sample of $>4$ Gyr stars with precise ages and clear rotational modulation is required to determine whether or not a break in the power-law near $1$ Gyr is justified.

\subsection{Implications for young exoplanet radii}

Exoplanet radii are measured from the depths of their transit light curves. Unocculted dark starspots slightly increase the depths of transit light curves, giving the appearance of slightly larger planets. In this section we quantify the extent of exoplanet radius amplification by dark starspots.

The \citet{Mandel2002} transit model for a uniform source is computed 
\begin{equation}
    F^e(t) = 1 - \lambda^e(t), \label{eqn:transit}
\end{equation}
where $\lambda^e$(t) is the fraction of the host star eclipsed. However, this is assuming the eclipsed body is not changing in brightness. If the host star is changing in brightness, Equation~\ref{eqn:transit} becomes
\begin{equation}
    F^e(t) = F_\star(t) - \lambda^e(t),
\end{equation}
where $F_\star(t)$ is the brightness of the star as a function of time relative to the unspotted stellar flux. In practice, observations of individual transits are often normalized by the out-of-transit flux immediately preceding and following the transit event, so what is observed is 
\begin{equation}
    F_{\rm obs}(t) \approx \frac{F_\star(t) - \lambda^e(t)}{F_\star(t)}.
\end{equation}
For a star with quadratic limb-darkening, for example, the maximum flux obscured during the transit event is  
\begin{equation}
\begin{split}
    \lambda^e_{\rm max} = \left(\frac{R_p}{R_\star}\right)^2 \times \\ \frac{1-u_1(1-\sqrt{1-b^2})-u_2 (1-\sqrt{1-b^2})^2}{1-\tfrac{u_1}{3}-\tfrac{u_2}{6}}, \label{eqn:lammax}
\end{split}
\end{equation}
where $b$ is the impact parameter, so the minimum observed flux during a transit event $F_{\rm obs, min}$ is 
\begin{equation}
    F_{\rm obs, min} \approx 1 - \frac{\lambda^e_{\rm max}}{F_\star(t)}.
\end{equation}
Since a spotted star has $F_\star(t) \leq 1$, it is clear that the transit depth $\delta_{\rm obs} = \lambda^e_{\rm max}/F_\star$ is amplified by a dimmer, more spotted surface. It is straightforward to compute $F_\star(t)$ for spotted stars with \fleck, so we can easily compute $\delta_{\rm obs}$ given parameters for the planet and starspots. 

\begin{figure}
    \centering
    \includegraphics[scale=0.8]{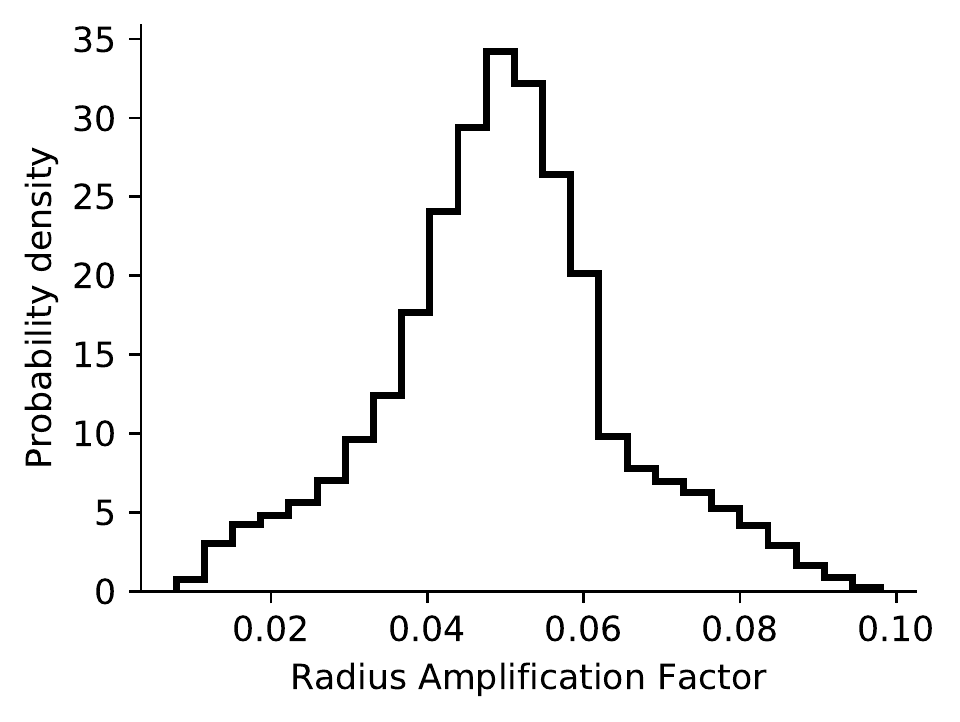}
    \caption{Apparent radius amplification due to transits at different points in the starspot modulation, for a very spotted star with $f_S = 0.12$ and a giant transiting exoplanet. The maximum contamination in the radius measurement due to starspots is about $10\%$ in this extreme case. }
    \label{fig:rad_amp}
\end{figure}

We simulate a star with three large spots of radius $\rm R_{spot} / R_\star = 0.4$, with stellar inclination $i_\star = 90^\circ$, and a planet with $\rm R_p/R_\star = 0.1$ at impact parameter $b=0$ with solar quadratic limb-darkening parameters. This is equivalent to the very large spot coverage $f_S = 0.12$. The results are shown in Figure~\ref{fig:rad_amp}. The transit depth can be amplified by as much as $10\%$ by the modulation due to starspots. This is large enough to be detected given typical observational uncertainties on the radii of small planets orbiting FGK stars, so the variation in transit depth as a function of time is likely an important factor for accurately measuring exoplanet radii orbiting stars with ages $\lesssim 20$ Myr. 

\section{Summary} \label{sec:conclusion}

We present photometric amplitudes of rotational modulation for 531 F, G, and K stars in six associations ranging in ages from 10 Myr to 4 Gyr. The age and rotation period are anti-correlated with the amplitudes of the light curves, which follows a Skumanich-like spin-down relation with age (Figures~\ref{fig:prot_smap} and \ref{fig:age_smap}). Using the rotational modulation model \fleck (Figure~\ref{fig:fleck}), along with an Approximate Bayesian Computation technique (Figure~\ref{fig:reproduce_smamps}), we estimate the spot coverage of stars as a function of age (Figure~\ref{fig:fs}), and find that spot coverage decays like $t^n$ where $t$ is the stellar age in Gyr and $n = -0.37 \pm 0.16$, compatible with a decay of spot coverage like the $t^{-1/2}$ relation discovered by \citet{Skumanich1972}.

We measured spot coverages of several planet-hosting stars with precise ages by modeling their rotational modulation (Figure~\ref{fig:bespoke}), and found good agreement between the measured and predicted spot coverages from the power-law relation (Figure~\ref{fig:fs_comparison} and Table~\ref{tab:pred_vs_measured}).

Based on the spot coverage estimates for young stars, we estimate that variations in the baseline flux of young FGK stars with ages $\lesssim 20$ Myr can cause apparently amplified exoplanet radii by up to $10\%$ (Figure~\ref{fig:rad_amp}).

Jupyter notebooks are available online\footnote{\url{https://github.com/bmorris3/birthmarks}} which can be used to reproduce this analysis.


\software{\texttt{astropy} \citep{Astropy2013, Astropy2018}, \texttt{ipython} \citep{ipython}, \texttt{numpy} \citep{VanDerWalt2011}, \texttt{scipy} \citep{scipy},  \texttt{matplotlib} \citep{matplotlib}, \texttt{scikit-learn} \citep{scikit-learn}, \texttt{corner} \citep{Foreman-Mackey2016}, \texttt{astroquery} \citep{Ginsburg2019}, \texttt{lightkurve} \citep{Lightkurve}, \texttt{photutils} \citep{photutils}, \texttt{healpy} \citep{healpy2005,healpy2019}, \texttt{emcee} \citep{Foreman-Mackey2013}} 

\facilities{Kepler, K2, TESS}

\acknowledgements

We are grateful for insightful conversations with Suzanne Hawley, Michele Bannister, David Fleming, Daniela Huppenkothen, Kevin Heng, Jason Curtis, Chloe Fisher, James Davenport and Erik Tollerud. We are also indebted to Thomas Affatigato, who first showed the author Praesepe through a telescope in 2006, and who planted other important seeds that made this work possible. This work has been carried out in the framework of the PlanetS National Centre of Competence in Research (NCCR) supported by the Swiss National Science Foundation (SNSF). This paper includes data collected by the Kepler mission. Funding for the Kepler mission is provided by the NASA Science Mission directorate. This paper includes data collected by the TESS mission. Funding for the TESS mission is provided by the NASA Explorer Program. This research has made use of NASA's Astrophysics Data System. This research has made use of the VizieR catalogue access tool, CDS, Strasbourg, France (DOI: 10.26093/cds/vizier). The original description of the VizieR service was published in A\&AS 143, 23. Some of the results in this paper have been derived using the healpy and HEALPix package.

\appendix

\section{Anderson-Darling Statistic} \label{sec:ad}

\begin{figure*}[h]
    \centering
    \includegraphics[scale=0.95]{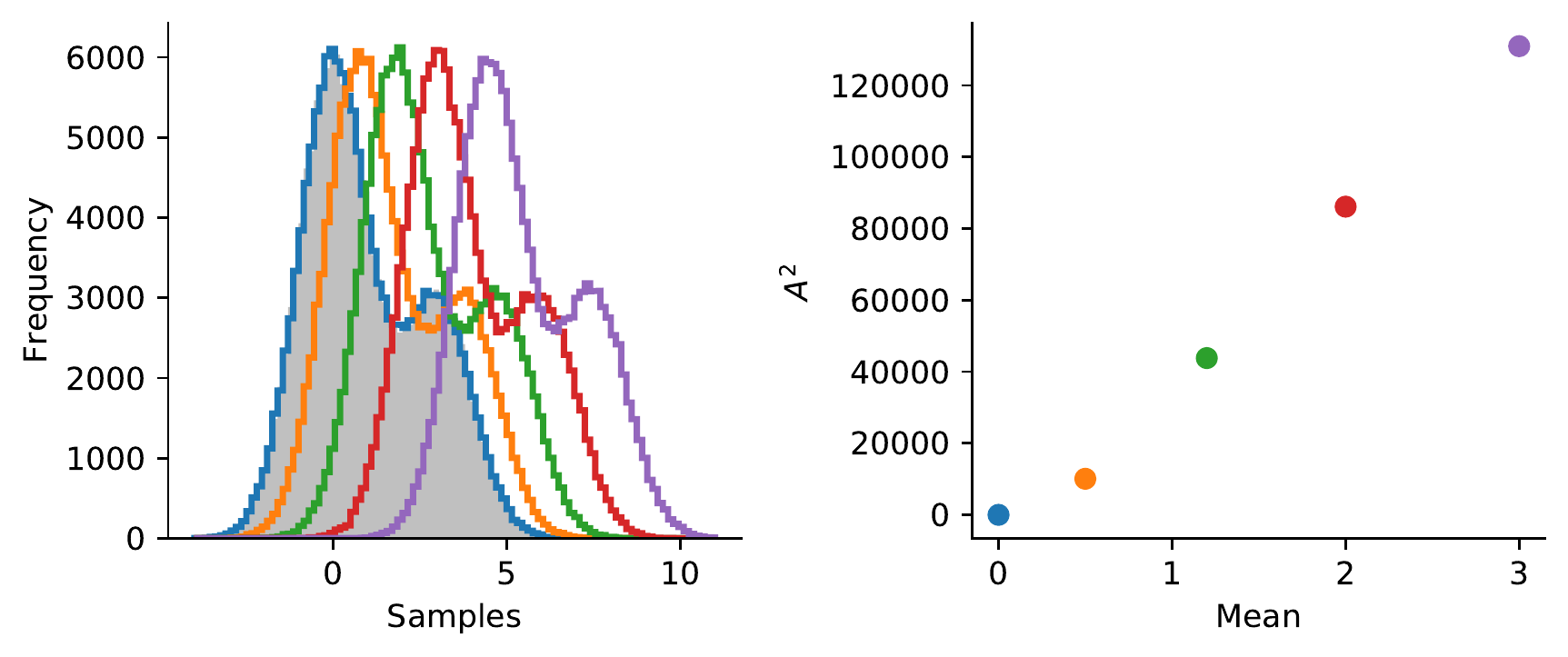}
    \caption{Demonstration of the behavior and dynamic range of the \texttt{scipy} implementation of the $k$-sample Anderson-Darling statistic $A^2$ when comparing pairs of bimodal distributions. The colored distributions on the left panel are compared with the gray-filled distribution, and the corresponding Anderson-Darling statistic is shown on the right panel with the same color. We chose a bimodal distribution for this example to illustrate that the Anderson-Darling statistic makes no assumption about the distributions being compared. The Anderson-Darling statistic for the sample with $\mu=0$ is $A^2\approx-0.6$.}
    \label{fig:ad}
\end{figure*}

The $k$-sample Anderson-Darling statistic is a non-parametric test of the null hypothesis that two groups of data are drawn from identical distributions. We use the \texttt{scipy} implementation of the $k$-sample Anderson-Darling statistic which varies from approximately -1.3 to $>10^5$, for distributions that are nearly identical and easily distinguishable, respectively. An exact description of the algorithm for computing the Anderson-Darling statistic can be found in \citet{Scholz1987}. A demonstration of the behavior and dynamic range of the Anderson-Darling statistic when comparing pairs of samples is given in Figure~\ref{fig:ad}.

\begin{thebibliography}{}
\expandafter\ifx\csname natexlab\endcsname\relax\def\natexlab#1{#1}\fi
\providecommand{\url}[1]{\href{#1}{#1}}
\providecommand{\dodoi}[1]{doi:~\href{http://doi.org/#1}{\nolinkurl{#1}}}
\providecommand{\doeprint}[1]{\href{http://ascl.net/#1}{\nolinkurl{http://ascl.net/#1}}}
\providecommand{\doarXiv}[1]{\href{https://arxiv.org/abs/#1}{\nolinkurl{https://arxiv.org/abs/#1}}}

\bibitem[{{Akeret} {et~al.}(2015){Akeret}, {Refregier}, {Amara}, {Seehars}, \&
  {Hasner}}]{Akeret2015}
{Akeret}, J., {Refregier}, A., {Amara}, A., {Seehars}, S., \& {Hasner}, C.
  2015, \jcap, 8, 043, \dodoi{10.1088/1475-7516/2015/08/043}

\bibitem[{{Alsubai} {et~al.}(2017){Alsubai}, {Mislis}, {Tsvetanov}, {Latham},
  {Bieryla}, {Buchhave}, {Esquerdo}, {Bramich}, {Pyrzas}, {Vilchez}, {Mancini},
  {Southworth}, {Evans}, {Henning}, \& {Ciceri}}]{Alsubai2017}
{Alsubai}, K., {Mislis}, D., {Tsvetanov}, Z.~I., {et~al.} 2017, \aj, 153, 200,
  \dodoi{10.3847/1538-3881/aa6340}

\bibitem[{Anderson \& Darling(1952)}]{anderson1952}
Anderson, T.~W., \& Darling, D.~A. 1952, Ann. Math. Statist., 23, 193,
  \dodoi{10.1214/aoms/1177729437}

\bibitem[{{Astropy Collaboration} {et~al.}(2013){Astropy Collaboration},
  {Robitaille}, {Tollerud}, {Greenfield}, {Droettboom}, {Bray}, {Aldcroft},
  {Davis}, {Ginsburg}, {Price-Whelan}, {Kerzendorf}, {Conley}, {Crighton},
  {Barbary}, {Muna}, {Ferguson}, {Grollier}, {Parikh}, {Nair}, {Unther},
  {Deil}, {Woillez}, {Conseil}, {Kramer}, {Turner}, {Singer}, {Fox}, {Weaver},
  {Zabalza}, {Edwards}, {Azalee Bostroem}, {Burke}, {Casey}, {Crawford},
  {Dencheva}, {Ely}, {Jenness}, {Labrie}, {Lian Lim}, {Pierfederici},
  {Pontzen}, {Ptak}, {Refsdal}, {Servillat}, \& {Streicher}}]{Astropy2013}
{Astropy Collaboration}, {Robitaille}, T.~P., {Tollerud}, E.~J., {et~al.} 2013,
  \aap, 558, A33, \dodoi{10.1051/0004-6361/201322068}

\bibitem[{{Astropy Collaboration} {et~al.}(2018){Astropy Collaboration},
  {Price-Whelan}, {Sip{\H o}cz}, {G{\"u}nther}, {Lim}, {Crawford}, {Conseil},
  {Shupe}, {Craig}, {Dencheva}, {Ginsburg}, {VanderPlas}, {Bradley},
  {P{\'e}rez-Su{\'a}rez}, {de Val-Borro}, {Aldcroft}, {Cruz}, {Robitaille},
  {Tollerud}, {Ardelean}, {Babej}, {Bach}, {Bachetti}, {Bakanov}, {Bamford},
  {Barentsen}, {Barmby}, {Baumbach}, {Berry}, {Biscani}, {Boquien}, {Bostroem},
  {Bouma}, {Brammer}, {Bray}, {Breytenbach}, {Buddelmeijer}, {Burke},
  {Calderone}, {Cano Rodr{\'{\i}}guez}, {Cara}, {Cardoso}, {Cheedella},
  {Copin}, {Corrales}, {Crichton}, {D'Avella}, {Deil}, {Depagne}, {Dietrich},
  {Donath}, {Droettboom}, {Earl}, {Erben}, {Fabbro}, {Ferreira}, {Finethy},
  {Fox}, {Garrison}, {Gibbons}, {Goldstein}, {Gommers}, {Greco}, {Greenfield},
  {Groener}, {Grollier}, {Hagen}, {Hirst}, {Homeier}, {Horton}, {Hosseinzadeh},
  {Hu}, {Hunkeler}, {Ivezi{\'c}}, {Jain}, {Jenness}, {Kanarek}, {Kendrew},
  {Kern}, {Kerzendorf}, {Khvalko}, {King}, {Kirkby}, {Kulkarni}, {Kumar},
  {Lee}, {Lenz}, {Littlefair}, {Ma}, {Macleod}, {Mastropietro}, {McCully},
  {Montagnac}, {Morris}, {Mueller}, {Mumford}, {Muna}, {Murphy}, {Nelson},
  {Nguyen}, {Ninan}, {N{\"o}the}, {Ogaz}, {Oh}, {Parejko}, {Parley}, {Pascual},
  {Patil}, {Patil}, {Plunkett}, {Prochaska}, {Rastogi}, {Reddy Janga},
  {Sabater}, {Sakurikar}, {Seifert}, {Sherbert}, {Sherwood-Taylor}, {Shih},
  {Sick}, {Silbiger}, {Singanamalla}, {Singer}, {Sladen}, {Sooley},
  {Sornarajah}, {Streicher}, {Teuben}, {Thomas}, {Tremblay}, {Turner},
  {Terr{\'o}n}, {van Kerkwijk}, {de la Vega}, {Watkins}, {Weaver}, {Whitmore},
  {Woillez}, {Zabalza}, \& {Astropy Contributors}}]{Astropy2018}
{Astropy Collaboration}, {Price-Whelan}, A.~M., {Sip{\H o}cz}, B.~M., {et~al.}
  2018, \aj, 156, 123, \dodoi{10.3847/1538-3881/aabc4f}

\bibitem[{{Baliunas} {et~al.}(1995){Baliunas}, {Donahue}, {Soon}, {Horne},
  {Frazer}, {Woodard-Eklund}, {Bradford}, {Rao}, {Wilson}, {Zhang}, {Bennett},
  {Briggs}, {Carroll}, {Duncan}, {Figueroa}, {Lanning}, {Misch}, {Mueller},
  {Noyes}, {Poppe}, {Porter}, {Robinson}, {Russell}, {Shelton}, {Soyumer},
  {Vaughan}, \& {Whitney}}]{Baliunas1995}
{Baliunas}, S.~L., {Donahue}, R.~A., {Soon}, W.~H., {et~al.} 1995, \apj, 438,
  269, \dodoi{10.1086/175072}

\bibitem[{{Barnes} {et~al.}(2016){Barnes}, {Weingrill}, {Fritzewski},
  {Strassmeier}, \& {Platais}}]{Barnes2016}
{Barnes}, S.~A., {Weingrill}, J., {Fritzewski}, D., {Strassmeier}, K.~G., \&
  {Platais}, I. 2016, \apj, 823, 16, \dodoi{10.3847/0004-637X/823/1/16}

\bibitem[{{Berdyugina}(2005)}]{Berdyugina2005}
{Berdyugina}, S.~V. 2005, Living Reviews in Solar Physics, 2, 8,
  \dodoi{10.12942/lrsp-2005-8}

\bibitem[{{Borucki} {et~al.}(2010){Borucki}, {Koch}, {Basri}, {Batalha},
  {Brown}, {Caldwell}, {Caldwell}, {Christensen-Dalsgaard}, {Cochran},
  {DeVore}, {Dunham}, {Dupree}, {Gautier}, {Geary}, {Gilliland}, {Gould},
  {Howell}, {Jenkins}, {Kondo}, {Latham}, {Marcy}, {Meibom}, {Kjeldsen},
  {Lissauer}, {Monet}, {Morrison}, {Sasselov}, {Tarter}, {Boss}, {Brownlee},
  {Owen}, {Buzasi}, {Charbonneau}, {Doyle}, {Fortney}, {Ford}, {Holman},
  {Seager}, {Steffen}, {Welsh}, {Rowe}, {Anderson}, {Buchhave}, {Ciardi},
  {Walkowicz}, {Sherry}, {Horch}, {Isaacson}, {Everett}, {Fischer}, {Torres},
  {Johnson}, {Endl}, {MacQueen}, {Bryson}, {Dotson}, {Haas}, {Kolodziejczak},
  {Van Cleve}, {Chandrasekaran}, {Twicken}, {Quintana}, {Clarke}, {Allen},
  {Li}, {Wu}, {Tenenbaum}, {Verner}, {Bruhweiler}, {Barnes}, \&
  {Prsa}}]{Borucki2010}
{Borucki}, W.~J., {Koch}, D., {Basri}, G., {et~al.} 2010, Science, 327, 977,
  \dodoi{10.1126/science.1185402}

\bibitem[{{Borucki} {et~al.}(2011){Borucki}, {Koch}, {Basri}, {Batalha},
  {Brown}, {Bryson}, {Caldwell}, {Christensen-Dalsgaard}, {Cochran}, {DeVore},
  {Dunham}, {Gautier}, {Geary}, {Gilliland}, {Gould}, {Howell}, {Jenkins},
  {Latham}, {Lissauer}, {Marcy}, {Rowe}, {Sasselov}, {Boss}, {Charbonneau},
  {Ciardi}, {Doyle}, {Dupree}, {Ford}, {Fortney}, {Holman}, {Seager},
  {Steffen}, {Tarter}, {Welsh}, {Allen}, {Buchhave}, {Christiansen}, {Clarke},
  {Das}, {D{\'e}sert}, {Endl}, {Fabrycky}, {Fressin}, {Haas}, {Horch},
  {Howard}, {Isaacson}, {Kjeldsen}, {Kolodziejczak}, {Kulesa}, {Li}, {Lucas},
  {Machalek}, {McCarthy}, {MacQueen}, {Meibom}, {Miquel}, {Prsa}, {Quinn},
  {Quintana}, {Ragozzine}, {Sherry}, {Shporer}, {Tenenbaum}, {Torres},
  {Twicken}, {Van Cleve}, {Walkowicz}, {Witteborn}, \& {Still}}]{Borucki2011}
{Borucki}, W.~J., {Koch}, D.~G., {Basri}, G., {et~al.} 2011, \apj, 736, 19,
  \dodoi{10.1088/0004-637X/736/1/19}

\bibitem[{Bradley {et~al.}(2016)Bradley, Sipocz, Robitaille, Tollerud,
  Vinícius, Deil, Barbary, Günther, Cara, Droettboom, Bostroem, Bray,
  Bratholm, Pickering, Craig, Barentsen, Pascual, adonath, Greco, Kerzendorf,
  StuartLittlefair, Ferreira, D'Eugenio, \& Weaver}]{photutils}
Bradley, L., Sipocz, B., Robitaille, T., {et~al.} 2016, astropy/photutils:
  v0.3, \dodoi{10.5281/zenodo.164986}

\bibitem[{{Bruno} {et~al.}(2018){Bruno}, {Lewis}, {Stevenson}, {Filippazzo},
  {Hill}, {Fraine}, {Wakeford}, {Deming}, {L{\'o}pez-Morales}, \&
  {Alam}}]{Bruno2018}
{Bruno}, G., {Lewis}, N.~K., {Stevenson}, K.~B., {et~al.} 2018, \aj, 156, 124,
  \dodoi{10.3847/1538-3881/aac6db}

\bibitem[{Bruno {et~al.}(2019)Bruno, Lewis, Alam, López-Morales, Barstow,
  Wakeford, Sing, Henry, Ballester, Bourrier, Buchhave, Cohen, Mikal-Evans,
  García Muñoz, Lavvas, \& Sanz-Forcada}]{Bruno2019}
Bruno, G., Lewis, N.~K., Alam, M.~K., {et~al.} 2019, Monthly Notices of the
  Royal Astronomical Society, 491, 5361, \dodoi{10.1093/mnras/stz3194}

\bibitem[{{Christiansen} {et~al.}(2012){Christiansen}, {Jenkins}, {Caldwell},
  {Burke}, {Tenenbaum}, {Seader}, {Thompson}, {Barclay}, {Clarke}, {Li},
  {Smith}, {Stumpe}, {Twicken}, \& {Van Cleve}}]{Christiansen2012}
{Christiansen}, J.~L., {Jenkins}, J.~M., {Caldwell}, D.~A., {et~al.} 2012,
  \pasp, 124, 1279, \dodoi{10.1086/668847}

\bibitem[{{Ciardi} {et~al.}(2018){Ciardi}, {Crossfield}, {Feinstein},
  {Schlieder}, {Petigura}, {David}, {Bristow}, {Patel}, {Arnold}, {Benneke},
  {Christiansen}, {Dressing}, {Fulton}, {Howard}, {Isaacson}, {Sinukoff}, \&
  {Thackeray}}]{Ciardi2018}
{Ciardi}, D.~R., {Crossfield}, I. J.~M., {Feinstein}, A.~D., {et~al.} 2018,
  \aj, 155, 10, \dodoi{10.3847/1538-3881/aa9921}

\bibitem[{{Corsaro} {et~al.}(2017){Corsaro}, {Lee}, {Garc{\'\i}a},
  {Hennebelle}, {Mathur}, {Beck}, {Mathis}, {Stello}, \&
  {Bouvier}}]{Corsaro2017}
{Corsaro}, E., {Lee}, Y.-N., {Garc{\'\i}a}, R.~A., {et~al.} 2017, Nature
  Astronomy, 1, 0064, \dodoi{10.1038/s41550-017-0064}

\bibitem[{{Curtis} {et~al.}(2019{\natexlab{a}}){Curtis}, {Ag{\"u}eros},
  {Douglas}, \& {Meibom}}]{Curtis2019}
{Curtis}, J.~L., {Ag{\"u}eros}, M.~A., {Douglas}, S.~T., \& {Meibom}, S.
  2019{\natexlab{a}}, \apj, 879, 49, \dodoi{10.3847/1538-4357/ab2393}

\bibitem[{{Curtis} {et~al.}(2019{\natexlab{b}}){Curtis}, {Ag{\"u}eros},
  {Mamajek}, {Wright}, \& {Cummings}}]{Curtis2019b}
{Curtis}, J.~L., {Ag{\"u}eros}, M.~A., {Mamajek}, E.~E., {Wright}, J.~T., \&
  {Cummings}, J.~D. 2019{\natexlab{b}}, \aj, 158, 77,
  \dodoi{10.3847/1538-3881/ab2899}

\bibitem[{{David} {et~al.}(2019{\natexlab{a}}){David}, {Petigura}, {Luger},
  {Foreman-Mackey}, {Livingston}, {Mamajek}, \& {Hillenbrand}}]{David2019}
{David}, T.~J., {Petigura}, E.~A., {Luger}, R., {et~al.} 2019{\natexlab{a}},
  \apjl, 885, L12, \dodoi{10.3847/2041-8213/ab4c99}

\bibitem[{{David} {et~al.}(2019{\natexlab{b}}){David}, {Cody}, {Hedges},
  {Mamajek}, {Hillenbrand}, {Ciardi}, {Beichman}, {Petigura}, {Fulton},
  {Isaacson}, {Howard}, {Gagn{\'e}}, {Saunders}, {Rebull}, {Stauffer},
  {Vasisht}, \& {Hinkley}}]{David2019b}
{David}, T.~J., {Cody}, A.~M., {Hedges}, C.~L., {et~al.} 2019{\natexlab{b}},
  \aj, 158, 79, \dodoi{10.3847/1538-3881/ab290f}

\bibitem[{{Douglas} {et~al.}(2017){Douglas}, {Ag{\"u}eros}, {Covey}, \&
  {Kraus}}]{Douglas2017}
{Douglas}, S.~T., {Ag{\"u}eros}, M.~A., {Covey}, K.~R., \& {Kraus}, A. 2017,
  \apj, 842, 83, \dodoi{10.3847/1538-4357/aa6e52}

\bibitem[{Dutta {et~al.}(2016)Dutta, Kaski, Lintusaari, Gutmann, \&
  Corander}]{Dutta2016}
Dutta, R., Kaski, S., Lintusaari, J., Gutmann, M.~U., \& Corander, J. 2016,
  Systematic Biology, 66, e66, \dodoi{10.1093/sysbio/syw077}

\bibitem[{{Foreman-Mackey}(2016)}]{Foreman-Mackey2016}
{Foreman-Mackey}, D. 2016, The Journal of Open Source Software, 1,
  \dodoi{10.21105/joss.00024}

\bibitem[{{Foreman-Mackey} {et~al.}(2013){Foreman-Mackey}, {Hogg}, {Lang}, \&
  {Goodman}}]{Foreman-Mackey2013}
{Foreman-Mackey}, D., {Hogg}, D.~W., {Lang}, D., \& {Goodman}, J. 2013, \pasp,
  125, 306, \dodoi{10.1086/670067}

\bibitem[{{Gagn{\'e}} {et~al.}(2018){Gagn{\'e}}, {Mamajek}, {Malo}, {Riedel},
  {Rodriguez}, {Lafreni{\`e}re}, {Faherty}, {Roy-Loubier}, {Pueyo}, {Robin}, \&
  {Doyon}}]{Gagne2018}
{Gagn{\'e}}, J., {Mamajek}, E.~E., {Malo}, L., {et~al.} 2018, \apj, 856, 23,
  \dodoi{10.3847/1538-4357/aaae09}

\bibitem[{{Gaia Collaboration} {et~al.}(2018){Gaia Collaboration}, {Brown},
  {Vallenari}, {Prusti}, {de Bruijne}, {Babusiaux}, \&
  {Bailer-Jones}}]{GaiaDR2}
{Gaia Collaboration}, {Brown}, A.~G.~A., {Vallenari}, A., {et~al.} 2018, ArXiv
  e-prints.
\newblock \doarXiv{1804.09365}

\bibitem[{{Giles} {et~al.}(2017){Giles}, {Collier Cameron}, \&
  {Haywood}}]{Giles2017}
{Giles}, H. A.~C., {Collier Cameron}, A., \& {Haywood}, R.~D. 2017, \mnras,
  472, 1618, \dodoi{10.1093/mnras/stx1931}

\bibitem[{{Ginsburg} {et~al.}(2019){Ginsburg}, {Sip{\H{o}}cz}, {Brasseur},
  {Cowperthwaite}, {Craig}, {Deil}, {Guillochon}, {Guzman}, {Liedtke}, {Lian
  Lim}, {Lockhart}, {Mommert}, {Morris}, {Norman}, {Parikh}, {Persson},
  {Robitaille}, {Segovia}, {Singer}, {Tollerud}, {de Val-Borro}, {Valtchanov},
  {Woillez}, {Astroquery Collaboration}, \& {a subset of astropy
  Collaboration}}]{Ginsburg2019}
{Ginsburg}, A., {Sip{\H{o}}cz}, B.~M., {Brasseur}, C.~E., {et~al.} 2019, \aj,
  157, 98, \dodoi{10.3847/1538-3881/aafc33}

\bibitem[{{Gonzalez}(2016{\natexlab{a}})}]{Gonzalez2016a}
{Gonzalez}, G. 2016{\natexlab{a}}, \mnras, 459, 1060,
  \dodoi{10.1093/mnras/stw700}

\bibitem[{{Gonzalez}(2016{\natexlab{b}})}]{Gonzalez2016b}
---. 2016{\natexlab{b}}, \mnras, 463, 3513, \dodoi{10.1093/mnras/stw2237}

\bibitem[{{G{\'o}rski} {et~al.}(2005){G{\'o}rski}, {Hivon}, {Banday},
  {Wandelt}, {Hansen}, {Reinecke}, \& {Bartelmann}}]{healpy2005}
{G{\'o}rski}, K.~M., {Hivon}, E., {Banday}, A.~J., {et~al.} 2005, \apj, 622,
  759, \dodoi{10.1086/427976}

\bibitem[{{Gully-Santiago} {et~al.}(2017){Gully-Santiago}, {Herczeg},
  {Czekala}, {Somers}, {Grankin}, {Covey}, {Donati}, {Alencar}, {Hussain},
  {Shappee}, {Mace}, {Lee}, {Holoien}, {Jose}, \& {Liu}}]{Gully2017}
{Gully-Santiago}, M.~A., {Herczeg}, G.~J., {Czekala}, I., {et~al.} 2017, \apj,
  836, 200, \dodoi{10.3847/1538-4357/836/2/200}

\bibitem[{{Hall}(2008)}]{Hall2008}
{Hall}, J.~C. 2008, Living Reviews in Solar Physics, 5, 2,
  \dodoi{10.12942/lrsp-2008-2}

\bibitem[{{H{\'e}brard} {et~al.}(2013){H{\'e}brard}, {Collier Cameron},
  {Brown}, {D{\'\i}az}, {Faedi}, {Smalley}, {Anderson}, {Armstrong}, {Barros},
  {Bento}, {Bouchy}, {Doyle}, {Enoch}, {G{\'o}mez Maqueo Chew}, {H{\'e}brard},
  {Hellier}, {Lendl}, {Lister}, {Maxted}, {McCormac}, {Moutou}, {Pollacco},
  {Queloz}, {Santerne}, {Skillen}, {Southworth}, {Tregloan-Reed}, {Triaud},
  {Udry}, {Vanhuysse}, {Watson}, {West}, \& {Wheatley}}]{Hebrard2013}
{H{\'e}brard}, G., {Collier Cameron}, A., {Brown}, D.~J.~A., {et~al.} 2013,
  \aap, 549, A134, \dodoi{10.1051/0004-6361/201220363}

\bibitem[{{Howard} {et~al.}(1984){Howard}, {Gilman}, \& {Gilman}}]{Howard1984}
{Howard}, R., {Gilman}, P.~I., \& {Gilman}, P.~A. 1984, \apj, 283, 373,
  \dodoi{10.1086/162315}

\bibitem[{{Howell} {et~al.}(2014){Howell}, {Sobeck}, {Haas}, {Still},
  {Barclay}, {Mullally}, {Troeltzsch}, {Aigrain}, {Bryson}, {Caldwell},
  {Chaplin}, {Cochran}, {Huber}, {Marcy}, {Miglio}, {Najita}, {Smith},
  {Twicken}, \& {Fortney}}]{Howell2014}
{Howell}, S.~B., {Sobeck}, C., {Haas}, M., {et~al.} 2014, \pasp, 126, 398,
  \dodoi{10.1086/676406}

\bibitem[{{Hunter}(2007)}]{matplotlib}
{Hunter}, J.~D. 2007, Computing in Science and Engineering, 9, 90,
  \dodoi{10.1109/MCSE.2007.55}

\bibitem[{{Jackson} \& {Jeffries}(2013)}]{Jackson2013}
{Jackson}, R.~J., \& {Jeffries}, R.~D. 2013, \mnras, 431, 1883,
  \dodoi{10.1093/mnras/stt304}

\bibitem[{{J{\"a}rvinen} {et~al.}(2018){J{\"a}rvinen}, {Strassmeier},
  {Carroll}, {Ilyin}, \& {Weber}}]{Jarvinen2018}
{J{\"a}rvinen}, S.~P., {Strassmeier}, K.~G., {Carroll}, T.~A., {Ilyin}, I., \&
  {Weber}, M. 2018, \aap, 620, A162, \dodoi{10.1051/0004-6361/201833496}

\bibitem[{Jones {et~al.}(2001)Jones, Oliphant, Peterson, {et~al.}}]{scipy}
Jones, E., Oliphant, T., Peterson, P., {et~al.} 2001, {SciPy}: Open source
  scientific tools for {Python}.
\newblock \url{http://www.scipy.org/}

\bibitem[{{Karoff} {et~al.}(2018){Karoff}, {Metcalfe}, {Santos}, {Montet},
  {Isaacson}, {Witzke}, {Shapiro}, {Mathur}, {Davies}, {Lund}, {Garcia},
  {Brun}, {Salabert}, {Avelino}, {van Saders}, {Egeland}, {Cunha}, {Campante},
  {Chaplin}, {Krivova}, {Solanki}, {Stritzinger}, \& {Knudsen}}]{Karoff2018}
{Karoff}, C., {Metcalfe}, T.~S., {Santos}, {\^A}. R.~G., {et~al.} 2018, \apj,
  852, 46, \dodoi{10.3847/1538-4357/aaa026}

\bibitem[{{Kenyon} \& {Hartmann}(1995)}]{Kenyon1995}
{Kenyon}, S.~J., \& {Hartmann}, L. 1995, \apjs, 101, 117,
  \dodoi{10.1086/192235}

\bibitem[{Kirk {et~al.}(2016)Kirk, Wheatley, Louden, Littlefair, Copperwheat,
  Armstrong, Marsh, \& Dhillon}]{Kirk2016}
Kirk, J., Wheatley, P.~J., Louden, T., {et~al.} 2016, Monthly Notices of the
  Royal Astronomical Society, 463, 2922, \dodoi{10.1093/mnras/stw2205}

\bibitem[{{Kovacs}(2018)}]{Kovacs2018}
{Kovacs}, G. 2018, \aap, 612, L2, \dodoi{10.1051/0004-6361/201731355}

\bibitem[{Lanza(2016)}]{Lanza2016}
Lanza, A.~F. 2016, Imaging Surface Spots from Space-Borne Photometry, ed. J.-P.
  Rozelot \& C.~Neiner (Cham: Springer International Publishing), 43--68,
  \dodoi{10.1007/978-3-319-24151-7_3}

\bibitem[{{Lightkurve Collaboration} {et~al.}(2018){Lightkurve Collaboration},
  {Cardoso}, {Hedges}, {Gully-Santiago}, {Saunders}, {Cody}, {Barclay}, {Hall},
  {Sagear}, {Turtelboom}, {Zhang}, {Tzanidakis}, {Mighell}, {Coughlin}, {Bell},
  {Berta-Thompson}, {Williams}, {Dotson}, \& {Barentsen}}]{Lightkurve}
{Lightkurve Collaboration}, {Cardoso}, J. V. d. M.~a., {Hedges}, C., {et~al.}
  2018, {Lightkurve: Kepler and TESS time series analysis in Python}.
\newblock \doeprint{1812.013}

\bibitem[{{Lomb}(1976)}]{Lomb1976}
{Lomb}, N.~R. 1976, \apss, 39, 447, \dodoi{10.1007/BF00648343}

\bibitem[{{Mandel} \& {Agol}(2002)}]{Mandel2002}
{Mandel}, K., \& {Agol}, E. 2002, \apjl, 580, L171, \dodoi{10.1086/345520}

\bibitem[{Mann {et~al.}(2017)Mann, Vanderburg, Rizzuto, Kraus, Berlind,
  Bieryla, Calkins, Esquerdo, Latham, Mace, Morris, Quinn, Sokal, \&
  Stefanik}]{Mann2017b}
Mann, A.~W., Vanderburg, A., Rizzuto, A.~C., {et~al.} 2017, The Astronomical
  Journal, 155, 4, \dodoi{10.3847/1538-3881/aa9791}

\bibitem[{{Mann} {et~al.}(2017){Mann}, {Gaidos}, {Vanderburg}, {Rizzuto},
  {Ansdell}, {Medina}, {Mace}, {Kraus}, \& {Sokal}}]{Mann2017}
{Mann}, A.~W., {Gaidos}, E., {Vanderburg}, A., {et~al.} 2017, \aj, 153, 64,
  \dodoi{10.1088/1361-6528/aa5276}

\bibitem[{{McQuillan} {et~al.}(2014){McQuillan}, {Mazeh}, \&
  {Aigrain}}]{McQuillan2014}
{McQuillan}, A., {Mazeh}, T., \& {Aigrain}, S. 2014, \apjs, 211, 24,
  \dodoi{10.1088/0067-0049/211/2/24}

\bibitem[{{Meibom} {et~al.}(2011){Meibom}, {Barnes}, {Latham}, {Batalha},
  {Borucki}, {Koch}, {Basri}, {Walkowicz}, {Janes}, {Jenkins}, {Van Cleve},
  {Haas}, {Bryson}, {Dupree}, {Furesz}, {Szentgyorgyi}, {Buchhave}, {Clarke},
  {Twicken}, \& {Quintana}}]{Meibom2011}
{Meibom}, S., {Barnes}, S.~A., {Latham}, D.~W., {et~al.} 2011, \apjl, 733, L9,
  \dodoi{10.1088/2041-8205/733/1/L9}

\bibitem[{{Meingast} {et~al.}(2019){Meingast}, {Alves}, \&
  {F{\"u}rnkranz}}]{Meingast2019}
{Meingast}, S., {Alves}, J., \& {F{\"u}rnkranz}, V. 2019, \aap, 622, L13,
  \dodoi{10.1051/0004-6361/201834950}

\bibitem[{{Morris} {et~al.}(2018){Morris}, {Agol}, {Davenport}, \&
  {Hawley}}]{Morris2018b}
{Morris}, B.~M., {Agol}, E., {Davenport}, J.~R.~A., \& {Hawley}, S.~L. 2018,
  \mnras, 476, 5408, \dodoi{10.1093/mnras/sty568}

\bibitem[{{Morris} {et~al.}(2019){Morris}, {Davenport}, {Giles}, {Hebb},
  {Hawley}, {Angus}, {Gilman}, \& {Agol}}]{Morris2019a}
{Morris}, B.~M., {Davenport}, J.~R.~A., {Giles}, H.~A.~C., {et~al.} 2019,
  \mnras, 484, 3244, \dodoi{10.1093/mnras/stz199}

\bibitem[{{Morris} {et~al.}(2017){Morris}, {Hebb}, {Davenport}, {Rohn}, \&
  {Hawley}}]{Morris2017a}
{Morris}, B.~M., {Hebb}, L., {Davenport}, J.~R.~A., {Rohn}, G., \& {Hawley},
  S.~L. 2017, \apj, 846, 99, \dodoi{10.3847/1538-4357/aa8555}

\bibitem[{{Newton} {et~al.}(2019){Newton}, {Mann}, {Tofflemire}, {Pearce},
  {Rizzuto}, {Vanderburg}, {Martinez}, {Wang}, {Ruffio}, {Kraus}, {Johnson},
  {Thao}, {Wood}, {Rampalli}, {Nielsen}, {Collins}, {Dragomir}, {Hellier},
  {Anderson}, {Barclay}, {Brown}, {Feiden}, {Hart}, {Isopi}, {Kielkopf},
  {Mallia}, {Nelson}, {Rodriguez}, {Stockdale}, {Waite}, {Wright}, {Lissauer},
  {Ricker}, {Vanderspek}, {Latham}, {Seager}, {Winn}, {Jenkins}, {Bouma},
  {Burke}, {Davies}, {Fausnaugh}, {Li}, {Morris}, {Mukai}, {Villase{\~n}or},
  {Villeneuva}, {De Rosa}, {Macintosh}, {Mengel}, {Okumura}, \&
  {Wittenmyer}}]{Newton2019}
{Newton}, E.~R., {Mann}, A.~W., {Tofflemire}, B.~M., {et~al.} 2019, \apjl, 880,
  L17, \dodoi{10.3847/2041-8213/ab2988}

\bibitem[{{{\"O}nehag} {et~al.}(2011){{\"O}nehag}, {Korn}, {Gustafsson},
  {Stempels}, \& {Vandenberg}}]{Onehag2011}
{{\"O}nehag}, A., {Korn}, A., {Gustafsson}, B., {Stempels}, E., \&
  {Vandenberg}, D.~A. 2011, \aap, 528, A85, \dodoi{10.1051/0004-6361/201015138}

\bibitem[{Pecaut \& Mamajek(2016)}]{Pecaut2016}
Pecaut, M.~J., \& Mamajek, E.~E. 2016, Monthly Notices of the Royal
  Astronomical Society, 461, 794, \dodoi{10.1093/mnras/stw1300}

\bibitem[{Pedregosa {et~al.}(2011)Pedregosa, Varoquaux, Gramfort, Michel,
  Thirion, Grisel, Blondel, Prettenhofer, Weiss, Dubourg, Vanderplas, Passos,
  Cournapeau, Brucher, Perrot, \& Duchesnay}]{scikit-learn}
Pedregosa, F., Varoquaux, G., Gramfort, A., {et~al.} 2011, Journal of Machine
  Learning Research, 12, 2825

\bibitem[{Perez \& Granger(2007)}]{ipython}
Perez, F., \& Granger, B.~E. 2007, Computing in Science and Engg., 9, 21,
  \dodoi{10.1109/MCSE.2007.53}

\bibitem[{{Pettersen} {et~al.}(1992){Pettersen}, {Hawley}, \&
  {Fisher}}]{Pettersen1992}
{Pettersen}, B.~R., {Hawley}, S.~L., \& {Fisher}, G.~H. 1992, \solphys, 142,
  197, \dodoi{10.1007/BF00156642}

\bibitem[{{Press} \& {Rybicki}(1989)}]{Press1989}
{Press}, W.~H., \& {Rybicki}, G.~B. 1989, \apj, 338, 277,
  \dodoi{10.1086/167197}

\bibitem[{{Ricker} {et~al.}(2014){Ricker}, {Winn}, {Vanderspek}, {Latham},
  {Bakos}, {Bean}, {Berta-Thompson}, {Brown}, {Buchhave}, {Butler}, {Butler},
  {Chaplin}, {Charbonneau}, {Christensen-Dalsgaard}, {Clampin}, {Deming},
  {Doty}, {De Lee}, {Dressing}, {Dunham}, {Endl}, {Fressin}, {Ge}, {Henning},
  {Holman}, {Howard}, {Ida}, {Jenkins}, {Jernigan}, {Johnson}, {Kaltenegger},
  {Kawai}, {Kjeldsen}, {Laughlin}, {Levine}, {Lin}, {Lissauer}, {MacQueen},
  {Marcy}, {McCullough}, {Morton}, {Narita}, {Paegert}, {Palle}, {Pepe},
  {Pepper}, {Quirrenbach}, {Rinehart}, {Sasselov}, {Sato}, {Seager},
  {Sozzetti}, {Stassun}, {Sullivan}, {Szentgyorgyi}, {Torres}, {Udry}, \&
  {Villasenor}}]{Ricker2014}
{Ricker}, G.~R., {Winn}, J.~N., {Vanderspek}, R., {et~al.} 2014, in Space
  Telescopes and Instrumentation 2014: Optical, Infrared, and Millimeter Wave,
  Vol. 9143, 914320, \dodoi{10.1117/12.2063489}

\bibitem[{{Schmitt} {et~al.}(2014){Schmitt}, {Agol}, {Deck}, {Rogers}, {Gazak},
  {Fischer}, {Wang}, {Holman}, {Jek}, {Margossian}, {Omohundro}, {Winarski},
  {Brewer}, {Giguere}, {Lintott}, {Lynn}, {Parrish}, {Schawinski}, {Schwamb},
  {Simpson}, \& {Smith}}]{Schmitt2014}
{Schmitt}, J.~R., {Agol}, E., {Deck}, K.~M., {et~al.} 2014, \apj, 795, 167,
  \dodoi{10.1088/0004-637X/795/2/167}

\bibitem[{Scholz \& Stephens(1987)}]{Scholz1987}
Scholz, F.~W., \& Stephens, M.~A. 1987, Journal of the American Statistical
  Association, 82, 918.
\newblock \url{http://www.jstor.org/stable/2288805}

\bibitem[{{Schrijver} {et~al.}(1989){Schrijver}, {Cote}, {Zwaan}, \&
  {Saar}}]{Schrijver1989}
{Schrijver}, C.~J., {Cote}, J., {Zwaan}, C., \& {Saar}, S.~H. 1989, \apj, 337,
  964, \dodoi{10.1086/167168}

\bibitem[{{Silva Aguirre} {et~al.}(2015){Silva Aguirre}, {Davies}, {Basu},
  {Christensen-Dalsgaard}, {Creevey}, {Metcalfe}, {Bedding}, {Casagrande},
  {Handberg}, {Lund}, {Nissen}, {Chaplin}, {Huber}, {Serenelli}, {Stello}, {Van
  Eylen}, {Campante}, {Elsworth}, {Gilliland}, {Hekker}, {Karoff}, {Kawaler},
  {Kjeldsen}, \& {Lundkvist}}]{Silva2015}
{Silva Aguirre}, V., {Davies}, G.~R., {Basu}, S., {et~al.} 2015, \mnras, 452,
  2127, \dodoi{10.1093/mnras/stv1388}

\bibitem[{{Sisson} {et~al.}(2018){Sisson}, {Fan}, \& {Beaumont}}]{Sisson2018}
{Sisson}, S.~A., {Fan}, Y., \& {Beaumont}, M.~A. 2018, arXiv e-prints,
  arXiv:1802.09720.
\newblock \doarXiv{1802.09720}

\bibitem[{{Skumanich}(1972)}]{Skumanich1972}
{Skumanich}, A. 1972, \apj, 171, 565, \dodoi{10.1086/151310}

\bibitem[{{Solanki}(2003)}]{Solanki2003}
{Solanki}, S.~K. 2003, \aapr, 11, 153, \dodoi{10.1007/s00159-003-0018-4}

\bibitem[{{Sonett} {et~al.}(1991){Sonett}, {Giampapa}, \&
  {Matthews}}]{Sonett1991}
{Sonett}, C.~P., {Giampapa}, M.~S., \& {Matthews}, M.~S. 1991, {The Sun in
  Time}

\bibitem[{{Strassmeier} \& {Rice}(1998)}]{Strassmeier1998}
{Strassmeier}, K.~G., \& {Rice}, J.~B. 1998, \aap, 330, 685

\bibitem[{{Sun} {et~al.}(2019){Sun}, {Ioannidis}, {Gu}, {Schmitt}, {Wang}, \&
  {Kouwenhoven}}]{Sun2019}
{Sun}, L., {Ioannidis}, P., {Gu}, S., {et~al.} 2019, \aap, 624, A15,
  \dodoi{10.1051/0004-6361/201834275}

\bibitem[{{Sunn{\aa}ker} {et~al.}(2013){Sunn{\aa}ker}, {Busetto}, {Numminen},
  {Corander}, {Foll}, \& {Dessimoz}}]{Sunnaker2013}
{Sunn{\aa}ker}, M., {Busetto}, A.~G., {Numminen}, E., {et~al.} 2013, PLoS
  Computational Biology, 9, e1002803, \dodoi{10.1371/journal.pcbi.1002803}

\bibitem[{{Van Der Walt} {et~al.}(2011){Van Der Walt}, {Colbert}, \&
  {Varoquaux}}]{VanDerWalt2011}
{Van Der Walt}, S., {Colbert}, S.~C., \& {Varoquaux}, G. 2011, ArXiv e-prints.
\newblock \doarXiv{1102.1523}

\bibitem[{Zonca {et~al.}(2019)Zonca, Singer, Lenz, Reinecke, Rosset, Hivon, \&
  Gorski}]{healpy2019}
Zonca, A., Singer, L., Lenz, D., {et~al.} 2019, Journal of Open Source
  Software, 4, 1298, \dodoi{10.21105/joss.01298}

\end{thebibliography}
\end{document}